\documentclass[preprintnumbers,amsmath,11pt,amssymb,floatfix,superscriptaddress,nofootinbib]{article}

\def\mytitle#1{\setcounter{equation}{0}
\setcounter{footnote}{0}
\begin{flushleft}\Large\textbf{#1}\end{flushleft}
\vspace{0.20cm}}
\def\myname#1{\leftline{{\large #1}}\vspace{-0.13cm}}
\def\myplace#1#2{\small\begin{flushleft}\textit{#1}\\
\texttt{#2}\end{flushleft}}

\def\myclassification#1{\small\noindent
Keywords : Dark Energy, Scale Factor, Redshift parametrization.
       #1\vspace{0.5cm}}
\usepackage{graphicx}
\usepackage{wrapfig}
\usepackage{lipsum} 
\usepackage{multirow}
\usepackage[utf8]{inputenc}
\usepackage{array}
\usepackage{mathtools}
\usepackage{makecell}
\usepackage{amsmath}
\usepackage{bigints}
\usepackage{amssymb}
\usepackage[mathscr]{euscript}
\usepackage{tipa}
\usepackage{cite}

\makeatletter
\newcommand{\vo}{\vec{o}\@ifnextchar{^}{\,}{}}
\makeatother

\DeclareFontFamily{OT1}{pzc}{}
\DeclareFontShape{OT1}{pzc}{m}{it}{<-> s * [0.900] pzcmi7t}{}
\DeclareMathAlphabet{\mathpzc}{OT1}{pzc}{m}{it}
\begin{document}
\mytitle{Posing Constraints on the Free Parameters of a New Model of Dark Energy EoS : Responses Through Cosmological Behaviours}

\myname{$Promila ~Biswas^{*}$\footnote{promilabiswas8@gmail.com}, $Parthajit~ Roy^{**}$\footnote{roy.parthajit@gmail.com} and  $Ritabrata~
Biswas^{*}$\footnote{biswas.ritabrata@gmail.com}}
\myplace{${}^{*}$Department of Mathematics, The University of Burdwan, Golapbag Academic Complex, City : Burdwan-713104, District : Purba Bardhaman, State : West Bengal, Country : India.\\ ${}^{**}$Department of Computer Science, The University of Burdwan, Golapbag Academic Complex, City : Burdwan-713104, District : Purba Bardhaman, State : West Bengal, Country : India.} {}
 
\begin{abstract}
Since the late 1990’s observations of type Ia Supernova, our universe is predicted to experience a late time cosmic acceleration. Theoretical support to this observation were intended to be built via proposition of a hypothetical fluid which staying inside the universe exerts negative pressure. Cosmologists have prescribed many candidates for this exotic fluid so far. In this alley, a popular method is to choose time dependent equation of state parameter $\omega = \frac{p}{\rho}$ and to parametrize it as a function of redshift. Again some common families of such parametrizations are constructed among which different members justify different properties of observed universe. Mainly, these were model dependent studies which comprise free parameters to be constrained by different observations. In this present article, a new expression for redshift parametrization is considered and we have constrain its free parameters for two Hubble parameter vs redshift data sets. These data sets are obtained depending on two basic methodologies known as different ages method and baryonic acoustic oscillation method. Different confidence contours for our model are located under the constraints of said data sets. Besides, different thermodynamic parameters related to the evolution of our universe are analysed. It is notified that our model indicates towards a delayed dark matter model which mimics EoS = $-1(\Lambda)$ phenomena at the present epoch. Deceleration parameters behaviour’s are studied. It is noticed that a possibility of future deceleration may occur for this new model. Outcomes for both the data sets are compared with each other. 
\end{abstract}

\myclassification{}

\section{Introduction}
Almost twenty years ago from now, late time cosmic acceleration was pointed out for the first time ever by the observations of two independent supernova collaborating teams (Riess, A.G. et al. 1998; Perlmutter, S. et al. 1999). This has speculated that the cosmic solution is comprised of time independent, spatially homogeneous hypothetical matter density and constant positive space curvature. This need of exotic matter/ energy density, at first, led us to the establishment of cosmological constant, $`\Lambda$' model with $\Omega_{m} \approx 0.3$, $\Omega_{\Lambda} \approx 0.7.$ This model is preferred alternative to the $\Omega_{m} = 1$ scenario. Albert Einstein (1917), to build the first cosmological model, introduced cosmological constant term $(\Lambda)$ to his field equations of general relativity (GR). The motivation was to  make the related cosmic solution static. One can treat ``Cosmological Constant" as a constant valued energy density of the vacuum (Zeldovich, Y.B. 1968). As the dynamic cosmological models (Friedmann, A. 1922;  Lemaitre, G. 1927) were noted and cosmic expansion was discovered (Hubble, E. 1929), this $\Lambda$ term appeared as an unnecessary one. Cosmic Microwave Background (CMB) evidence for spatially flat universe (Bernardis, P. de et al. 2000; Hanany, S. et al. 2000), i.e., the proposition for $\Omega_{tot} \approx 1$ was declared soon after. With $\Omega_{m} << 1$ and $\Omega_{\Lambda} = 0$, this eliminated the free expansion theory. Scale factor $a(t)$, which is governed by GR, grows at an accelerating rate if the pressure $p < -\frac{1}{3}\rho$. Another popular methodology to explain the late time cosmic acceleration was introduced where a hypothetical fluid/ energy component is assumed to exert negative pressure staying at the right hand side of Einstein’s field equations. Names of different candidates for such exotic matter are coined as quintessence, phantom or dark energy (DE). Value of the equation of state $\omega = \frac{p}{\rho}$ is taken to be negative (Peebles,  P. J. E. \& Ratra, B. 2003; Biswas, P. \& Biswas, R. 2018a). Most natural existence of such energy density can be simply obtained if the cosmological constant term is reintroduced into field equations. Cosmological upper bound $(\rho \lesssim 10^{71}GeV^4)$ is satisfied by more than the order of hundreds if DE interpolation of $\Lambda$ is considered (Biswas, P. \& Biswas, R. 2018a). Besides this $\Lambda$CDM model, some $\omega$CDM models are also famous ($\omega$ is already mentioned DE EoS parameter). Huge discrepancies between observation and theory associated with the vacuum energy density enforces us to construct a better model. Time varying cosmological constant model (Sahni, V. \& Starobinsky, A. A. 2000), irreversible process of cosmological matter creation (Ozer, M. \& Taha, M. O. 1986), Chaplygin gas family (Biswas, P. \& Biswas, R. 2019a), redshift parametrization of the EoS parameters (Biswas, R. \& Debnath, U. 2013; Biswas, P. \& Biswas R. 2019b) etc are familiar ones. Scalar field $\phi$ with potential $V(\phi)$ (Ratra, B. \& Peebles,  P. J. E. 1988) is proposed as another model. In the limit $\frac{1}{2} {\dot{\phi}}^2 << |V(\phi)|$, this model acts like the cosmological constant. 

$H(z)= \frac{1}{a(t)} \frac{da(t)}{dt}$, the expansion rate of the universe is governed by the Friedmann equation. DE density can be written as a function of EoS $\omega(z)$ as

$$\frac{{\rho_{DE}}_{(z)}}{{\rho_{{DE}_0}}} =\frac{\Omega_{DE}}{\Omega_{{DE}_0}}= exp \left[3\int_0^z \bigg\{\frac{1+\omega(z')}{1+z'}\bigg\}dz'\right]~~~~~~~~,$$

Where $\rho_{{DE}_0}=\rho_{DE}(z=0)$ and $\Omega_{{DE}_0}=\Omega_{DE}(z=0)$. In GR based linear perturbation theory, the pressureless matter's density contrast $\delta(\vec{x},t) \equiv \frac{\rho (\vec{x},t)}{\rho(t)} -1$ grows in proportion to linear growth function $G(t)$. This is followed by the differential equation
$$\ddot{G} + 2H(z) \dot{G} - \frac{3}{2}\Omega_m H_0^2 (1+z)^3 G = 0~~~~~~~~~,$$ where ``." denotes differentiation with respect to $t$.

Upto a good approximation, logarithmic derivative of $G(z)$ is $$f(z) \equiv -\frac{d\{ln G\}}{d \{ln (1+z)\}} \approx \left[\Omega_m (1+z)^3 \frac{H_0^2}{H^2(z)}\right]^{\gamma}~~~~~,$$
where relevant values of cosmological parameters (Linder,  E.V. 2005) supports $\gamma \approx 0.55$.

Conventions exist there to phase constrain $\omega(z)$ in terms of linear evaluation model, $\omega(a) = \omega_0 + \omega_a (1-a) = \omega_p +  \omega_a (a_p-a)$, where $a = \frac{1}{1+z}$, $\omega_0$ is the value of $\omega$ at $z=0$ and $\omega_p$ is the value of $\omega$ at pivot redshift $z_p$ = $a_p^{-1}-1$.

More general set of cosmological parameters is constructed. Some necessary parameters are :- 
\begin{itemize}

\item[(i)] Dimensionless Hubble's parameter $h = \frac{H_0}{100}Kms^{-1}Mpc^{-1}$ which determines the present day value of critical density and overall scaling of distances inferred from redshifts.
\item[(ii)]$\Omega_m$ and $\Omega_{tot}$ affect expansion history and ``distance-redshift" relations 
\item[(iii)] The comoving distances through that the pressure wave is able to propagate between $t=0$ and recombination, the sound horizon $r_s = \int_0^{t_{rec}} \frac{c_s (t)}{a(t)} dt$, $c_s$ is the sound speed through the exotic fluid, determines the physical scale for the Acoustic peaks in the CMB (Pan, Z. 2016) and BAO feature in low redshift matter clustering (Bassett, B. A. \& Hlozek, R. 2010).
\item[(iv)] The quantity $\sigma_8(z)$ represents the amplitude of matter fluctuations and scales the overall amplitude of growth measurements. These may include weak density or redshift-space distortions. Time dependent modelling of DE EoS has a particular alley named as redshift parametrization which can not be obtained from the scale field dynamics. This is because of the fact that  these parametrizations are neither limited functions nor they lie in the interval defined by $\omega = \frac{\frac{\dot{\phi}^2}{2} - V(\phi)}{\frac{\dot{\phi}^2}{2} + V(\phi)}$ with scalar field $\phi$ and field potential $V(\phi)$. Two familiar series of redshift dependent EoS are found, viz, 
\end{itemize}

\textbf{Series I :} $\omega(z) = \omega_0 + \omega_1 \left(1-\frac{1}{1+z} \right)^n$ and 
\textbf{Series II :} $\omega(z) = \omega_0 + \omega_1 \frac{z}{(1+z)^n}$, ${\bf n \in \mathbb{N}}$.

Some other families such as Barboza-Alcaniz (2008), Efstathiou parametrization (Efstathiou, G. 1999; Silva, R. et al 2007), ASSS parametrization (Alam, U. et al 2004a, 2004b), Hannestad M{\"o}rtsell Parametrization (Hannestad, S. \& Mörtsell, E. 2004), Lee Parametrization (Lee, S. 2005), Feng Shen Li Li (FSLL) Parametrization (Feng, C., J. et al 2012), Polynomial Parametrization (Weller, J. \& Albrecht, A. 2002; Sendra, I. \& Lazkoz, R. 2012) can be found in literature.

In this present article, our motive is to propose a new EoS to replicate $\omega(z) = -1$ epoch at $z=0$. We will try to constrain the free parameters of the model under the observational data. While proposing the DE EoS we take care that it does not fall in any of Series I or Series II. The model has a completely new structure depending on redshift $z$. We will try to show whether our model generates $\omega(z)=-1$ epoch in the neighbourhood of $z=0$ or not. Tendency to occur a future deceleration will also be examined. We wish to study the nature of fractional dimensionless density parameters for our model. We plan to study the behaviour of deceleration parameter $q(z)$ for the proposed model and whether phase transition-(s) from deceleration to acceleration or the converse take(s) place or not.

The article is organised as follows : In section 2, we construct the cosmological model for our newly proposed parameterization. In section 3, we constrain the model under the data sets obtained with the help of differential ages method and Baryonic Acoustic Oscillation method. Section 4 comprises studies of different cosmological parameters related to our model. Finally, we briefly discuss our findings and conclude in the last section.

\section{Mathematical Construction of a New Kind of Redshift Parametrization}
Einstein's field equation in homogeneous and isotropic Friedmann–Lemaître–Robertson–Walker (FLRW) space time (with flat section) is written as 
\begin{equation}\label{Einstein_field_equations_for_FLRW_space_time_I}
3\frac{\dot{a}^2}{a^2} = \rho_{dm} + \frac{1}{2}{\dot{c\phi}}^2 + V(c\phi) = \rho_{dm} + \rho_{c\phi} + \rho_{rad}
\end{equation}
and
\begin{equation}\label{Einstein_field_equations_for_FLRW_space_time_II}
2\frac{\ddot{a}}{a} + \frac{\dot{a}^2}{a^2} = -\frac{1}{2}{\dot{c\phi}}^2 + V(c\phi) = -p_{c\phi}~~~~~~~~~~~~~,
\end{equation}
with Planck's units $8\pi G=c=1$, as stated earlier, $c\phi$ is the scalar field in natural units, $\rho_{c\phi}$, $p_{c\phi}$ and $V(c\phi)$ are the scalar field's density, pressure and potential respectively. $\rho_{dm}$ being the matter density.

Energy density and pressure of scalar field, $\rho_{c\phi}$ and $p_{c\phi}$ should have the structure as 
\begin{equation}\label{cphidensity3}
\rho_{c\phi} = \frac{1}{2}\dot{c\phi}^2 + V(c\phi)~~~~~~and~~~~~~p_{c\phi} = \frac{1}{2}\dot{c\phi}^2 - V(c\phi)~~~~~~~~~~.
\end{equation}
We will write non interactive DE-dark matter conservation equations
\begin{equation}\label{scalar_field_equation}
\dot{\rho_{c\phi}} + 3H(\rho_{c\phi} + p_{c\phi}) = 0~~~~~~~~for~energy
\end{equation}
and
\begin{equation}\label{matter_field}
\dot{\rho_{dm}} + 3H \rho_{dm} = 0 \Rightarrow \rho_{dm} = \rho_{dm0}a^{-3}~~~~~for~matter,
\end{equation}
where $\rho_{dm0}$ denotes the current time value of the matter field's energy density. Equation (\ref{scalar_field_equation}) can be rewritten as
\begin{equation}
\omega_{c\phi} = \frac{p_{c\phi}}{\rho_{c\phi}} = -1- \frac{a}{3 \rho_{c\phi}}\frac{d \rho_{c\phi}}{da}~~~~~~~~~~~~~~~~~~.
\end{equation}

Among the above equations (\ref{Einstein_field_equations_for_FLRW_space_time_I}), (\ref{Einstein_field_equations_for_FLRW_space_time_II}), (\ref{scalar_field_equation}) and (\ref{matter_field}), only three equations are likely to be independent to each other. Bianchi identities can show the derivation of the fourth equation. So, we are going to solve four independent variables. Without additional input, it is impossible to find an exact solution. We propose an ansatz for the functional form of $\rho_{c\phi}$ as 
\begin{equation}\label{new ansatz}
\frac{1}{\rho_{c\phi}}\frac{d \rho_{c\phi}}{d a} = -3 \bigg[\frac{\lambda_1}{1+a k_1} + \frac{\lambda_2 (1-a)}{(1+a k_2)^2}\bigg]~~~~~~~~~~,
\end{equation}
where $\lambda_1$, $\lambda_2$, $k_1$ and $k_2$ are constants\footnote{\bf{We propose this ansatz as we need four different equations to obtain an exact solution of four parameters $H$, $\rho_m$, $\phi$ and $V(\phi)$, by solving necessarily four equations of which only three we had. While proposing equation \eqref{new ansatz}, we have tried to increase a degree in $\frac{1}{1+z}$ from the CPL parametrization (Chevallier, M., Polarski, D. 2001; Linder, E.V. 2003) retaining the original terms. This might be treated as a modification of JBP parameterization (Jassal, H. K., Bagla, J. S., Padmanabhan, T. 2005) only by changing the constant of the denominator by parameters. We were careful to reproduce the $\Lambda = -1$ era at the present time epoch. A particular case of our assumption will lead to a EoS $\omega(a)=\omega_0+(\omega_1+\omega_2)a-\omega_2 a^2$ $\sim \omega_{\Lambda}+\omega_{f}a+\omega_{s}a^2$, where $\omega_{\Lambda}=-1$, a constant, $\omega_{f}$ and $\omega_{s}$ are two free parameters to be constrained by observation. The motivation is to examine how good our model fits with observational data and thermodynamic properties of the universe. This might be treated to reside in the combination of the first three components of the parameterization $\sum\limits^2_{n=0} \left(1-\frac{a}{a_0}\right)^n$. }}.

Integrating, we get
\begin{equation}
\rho_{c\phi} = A\frac{\left(1+a k_2\right)^{\frac{3 \lambda_2}{k_2^2}}}{\left(1+a k_1\right)^{\frac{3 \lambda_1}{k_1}}} exp \bigg\{\frac{3 \lambda_2 (1+k_2)}{k_2^2 (1+a k_2)}\bigg\}~~~~~~~~,
\end{equation}

where $A = \rho_{c\phi 0}\frac{\left(1+k_1\right)^{\frac{3 \lambda_1}{k_1}}}{\left(1+k_2\right)^{\frac{3 \lambda_2}{k_2^2}}} exp \bigg\{-\frac{3 \lambda_2}{k_2^2}\bigg\}$ and $\rho_{c\phi 0}$ is the present time (at $z = 0$) value of the scalar field density. We observe that the density depends on three distinct functions of $a$. If we make $\lambda_2 = 0$ and $k_1 = 1$, we see the solution will take a simple power law evolution of $\rho_{c\phi}(\sim a^{-\lambda})$, considered in many cosmological studies (Copeland, E. J. 2006). Equations (7) and (8) together give us the EoS parameter $\omega_{c\phi}$ as a function of redshift ($z = \frac{1}{a} - 1$) as
\begin{equation}\label{EoSPB}
\omega_{c\phi(z)} = -1 + \frac{\lambda_1}{(1 + k_1) + z} + \frac{\lambda_2 z}{\{(1 + k_2) + z\}^2}~~~~~~~~~~~~~~~.
\end{equation}
This equation even can be treated to be same of
\begin{equation}\label{newEoSPB}
\omega_{c\phi(z)} = \omega_0 + \frac{\omega_1}{\omega_2 + z} + \frac{\omega_3 z}{(\omega_4 + z)^2}~~~~~~~~~~~~~~~~~~.
\end{equation}
In the next section, we will constrain two of the free parameters of our model, namely, $\lambda_1$ \& $\lambda_2$ by the help of Hubble's parameter vs redshift data.

\section{Constraining the Free Parameters for DA and BAO method :}
Equation (10) depicts a new construction of DE EoS parameter. From (\ref{Einstein_field_equations_for_FLRW_space_time_I}), (\ref{matter_field}) and (9) we have

\begin{equation}\label{Obtained_Hubbles_parameter}
H^2 = H_0^2 \bigg[\Omega_{rad0}a^{-4} + \Omega_{dm0} a^{-3} + \Omega_{c\phi0}  \beta \frac{\left(1+a k_2\right)^{\frac{3 \lambda_2}{k_2^2}}}{\left(1+a k_1\right)^{\frac{3 \lambda_1}{k_1}}} exp\bigg\{\frac{3 \lambda_2 (1+k_2)}{k_2^2 (1 + a k_2)}\bigg\}\bigg]~~~~~~~~,
\end{equation}

where $\beta = \frac{\left(1+k_1\right)^{\frac{3 \lambda_1}{k_1}}}{\left(1+k_2\right)^{\frac{3 \lambda_2}{k_2^2}}} exp\{-\frac{3 \lambda_2}{k_2^2}\}$ is a constant and $\Omega_{dm0} = \frac{\rho_{dm0}}{3 H_0^2}$, $\Omega_{rad0} =  \frac{\rho_{rad0}}{3 H_0^2} $ and $\Omega_{c\phi 0} = \frac{\rho_{c\phi 0}}{3 H_0^2} = 1 - \Omega_{rad0} - \Omega_{dm0}$
represent the current values of the dimensionless density parameters for the matter, radiation and the scalar field respectively. Now we will proceed for constraining the parametrization's free parameters with Hubble parameter vs redshift data. For this we must enlist the data first and mention the methods for collecting the data.

While we look at the sky, if the celestial distance through which we see some objects is shorter one, Cepheid variables are used as standard candles. When the turn comes for looking at distant galaxies, type Ia Supernova explosions (SNeIa) are taken as standard candles (Riess, A.G. et al. 1998; Perlmutter, S. et al. 1999). Fluctuations in visible baryonic matter’s density which is caused by acoustic density waves in primordial plasma of the early universe is used as Standard rulers (Cole, S. et al. 2005; Eisenstein, D. J. et al. 2005). An acoustic wave can travel through primordial plasma until it is cooled to the point where it turns to neutral atoms and the maximum distance traversed by such a way is chosen to be the length of standard ruler. This oscillation’s name is coined as baryon acoustic oscillation (BAO). Since the last twenty two years, studies of cosmic microwave background (CMB) (remnant electromagnetic radiation came out of early Big Bang cosmology) have enriched the understanding of expanding universe along with the already mentioned candles and rulers. These methodologies, however, do not directly constrain the Hubble's parameter. Another independent methodology is named as ``cosmic chronometer approach" constructed to constrain the history of the universe's expansion (Jimenez, R. \& Loeb, A. 2002;  Valent, A. G. \& Amendola, L. 2018).

Moresco, M. et al. (2012) have analysed about $\approx$ 11,000 massive and passive galaxies and enlisted $H(z)$ value in the range of $0.15<z<1.1$ after eight measurements with accuracy of $5-12\%$. When $z<0.3$, we can find the most accurate constraints. Some authors (Moresco, M. et al 2012b; Wang, X. et al. 2012;  Zhao, G. -B.et al. 2012; Riemer-Sorensen, S. et al.  2013) have comparatively discussed the standard probe's (like BAO and SNeIa) and cosmic chronometer. In the works of Moresco, M. et al. (2016), some more Hubble parameter values are enlisted in the range of $0.35<z<0.5$.

A star's age can be calculated when we can analyse the spectra releasing out of it. We can take accumulation of stars, i.e., galaxies so that we can point out the ages. Thus we can observe a clock's behaviour and this ``so called clock" can be observed in archival data (Jimenez, R. et al 2003), GDDS- Gemini Deep Deep Survey (McCarthy, P. J. et al 2004).  We use BAO signature density auto-correlation function in size method as the ``standard rod". We believe that the entire theory is basically standing on the formation of a single ``burst" (Jimenez, R. et al 1999) of almost all stars of a galaxy.
 
We can notice more sensitive results for $\omega(z)$ in the differential ages (DA) case. In DA case, we must believe a clock and the dates of which may vary with the universe's age according to redshift. Spectroscopic dating of galaxy's ages provides this clock. We can write $\frac{\bigtriangleup z}{\bigtriangleup t}$ instead of $\frac{dz}{dt}$ based on $\bigtriangleup t$ and $\bigtriangleup z$, where $\bigtriangleup z$ is a small redshift interval and $\bigtriangleup t$ is the age difference measurement between two passively reappear galaxies that created at the same time. So it is a more genuine method rather than age determination for galaxies (Dunlop, J. et al. 1996; Alcaniz, J. S. \& Lima, J. A. S. 2001; Stockton, A. 2001). The study of globular clusters, absolute stellar ages are more absorbent to well connected ages. Moreover, we can find the lower limit to the age of the universe with respect to absolute galaxy’s ages and sort weak limitations of $\omega(z)$.

We can write Hubble parameter as $$dt = -\frac{(1+z)^{-1}}{H(z)} dz~~~~~~.$$ Applying this for old galaxies, one can obtain the value of $H_0$, i.e., using this method to the elliptical galaxies (in the local universe), we can evaluate the current value of Hubble constant.

Now we will form two tables for two types of $H(z)-z$ data sets and the corresponding error terms. First 46 data sets are provided by DA method and next, 26 data sets are provided by the standard ruler method, i.e., BAO method. 
\newpage
\begin{center}
Table-I : Hubble parameter $H(z)$ with redshift and errors $\sigma_H$ from DA method
\end{center}
\begin{minipage}{.5\linewidth}
	\centering
	\begin{tabular}{ ||>{\centering\arraybackslash}m{0.4cm}|>{\centering\arraybackslash}m{1cm}|>{\centering\arraybackslash}m{1cm}|>{\centering\arraybackslash}m{1cm}|>{\centering\arraybackslash}m{3.3cm}|| }
		\hline
		Sl No. & z & $H(z)$ & $\sigma(z)$ & Reference \\ 
		\hline
		1 & 0 & 67.77 & 1.30 & Macaulay, E. et al. 2018 \\
		\hline
		2 & 0.07 & 69 & 19.6 & Zhang, C. et al. 2014 \\ 
		\hline
		3 & 0.09 & 69 & 12 & Simon, J. et al. 2005  \\
		\hline
		4 & 0.1 & 69 & 12 & Stern, D. et al. 2010  \\ 
		\hline
		5 & 0.12 & 68.6 & 26.2 & Zhang, C. et al. 2014 \\
		\hline
		6 & 0.17 & 83 & 8 & Stern, D. et al. 2010 \\
		\hline
		7 & 0.179 & 75 & 4 & Moresco, M. et al. 2012a \\ 
		\hline
		8 & 0.1993 & 75 & 5 &  Moresco, M. et al. 2012a; Valent, A. G. \& Amendola, L.2018\\
		\hline
		9 & 0.2 & 72.9 & 29.6 & Zhang, C. et al. 2014  \\
		\hline
		10 & 0.24 & 79.7 & 2.7 & Gazta$\tilde{n}aga$,  E. et al. 2009\\
		\hline
		11 & 0.27 & 77 & 14 & Stern, D. et al. 2010 \\
		\hline
		12 & 0.28 & 88.8 & 36.6 & Zhang, C. et al. 2014 \\
		\hline
		13 & 0.35 & 82.7 & 8.4 & Chuang, C. H. \& Wang, Y. 2013 \\
		\hline
		14 & 0.352 & 83 & 14 & Moresco, M. 2015 \\
		\hline
		15 & 0.38 & 81.5 & 1.9 & Alam, S., et al. 2016\\
		\hline
		16 & 0.3802 & 83 & 13.5 & Moresco, M. et al. 2016 \\
		\hline
		17 & 0.4 & 95 & 17 & Simon, J. et al. 2005 \\
		\hline
		18 & 0.4004 & 77 & 10.2 & Moresco, M. et al. 2016  \\
		\hline
		19 & 0.4247 & 87.1 & 11.2 & Moresco, M. et al. 2016 \\
		\hline
		20 & 0.43 & 86.5 & 3.7 & Gazta$\tilde{n}aga$,  E. et al. 2009\\
		\hline
		21 & 0.44 & 82.6 & 7.8 & Blake, C. et al. 2012 \\
		\hline
		22 & 0.44497 & 92.8 & 12.9 & Moresco, M. et al. 2016 \\
		\hline
		23 & 0.47 & 89 & 49.6 &  Ratsimbazafy, A. L. et al. 2017; Valent, A. G. \& Amendola, L. 2018\\
		\hline 
			\end{tabular}
\end{minipage}
\begin{minipage}{.5\linewidth}
	\centering
	\begin{tabular}{ ||>{\centering\arraybackslash}m{0.4cm}|>{\centering\arraybackslash}m{1cm}|>{\centering\arraybackslash}m{1cm}|>{\centering\arraybackslash}m{1cm}|>{\centering\arraybackslash}m{3.12cm}|| }
		\hline
		Sl No. & z & $H(z)$ & $\sigma(z)$ & Reference \\ [0.5ex] 
		\hline
		24 & 0.4783 & 80.9 & 9 & Moresco, M. et al. 2016\\
		\hline
		25 & 0.48 & 97 & 60 & Stern, D. et al. 2010  \\
		\hline
		26 & 0.51 & 90.4 & 1.9 &  Alam, S., et al. 2016\\
		\hline
		27 & 0.57 & 96.8 & 3.4 & Anderson, L. et al. 2014  \\
		\hline
		28 & 0.593 & 104 & 13  & Moresco, M. et al. 2012a\\
		\hline 
		29 & 0.6 & 87.9 & 6.1 & Blake, C. et al. 2012\\
		\hline
		30 & 0.61 & 97.3 & 2.1 & Alam, S., et al. 2016\\
		\hline
		31 & 0.68 & 92 & 8 & Moresco, M. et al. 2012a\\
		\hline 
		32 & 0.73 & 97.3 & 7 & Blake, C. et al. 2012\\
		\hline
		33 & 0.781 & 105 & 12 & Moresco, M. et al. 2012a\\
		\hline
		34 & 0.875 & 125 & 17 & Moresco, M. et al. 2012a\\
		\hline
		35 & 0.88 &  90 & 40 & Stern, D. et al. 2010 \\
		\hline
		36 & 0.9 & 117 & 23 & Stern, D. et al. 2010 \\
		\hline
		37 & 1.037 & 154 & 20 & Moresco, M. et al. 2012a\\
		\hline
		38 & 1.3 & 168 & 17 & Stern, D. et al. 2010 \\
		\hline
		39 & 1.363 & 160 & 33.6 & Moresco, M. 2015\\
		\hline
		40 & 1.43 & 177 & 18 & Stern, D. et al. 2010 \\
		\hline
		41 & 1.53 & 140 & 14 & Stern, D. et al. 2010 \\
		\hline
		42 & 1.75 & 202 & 40 & Stern, D. et al. 2010 \\
		\hline
		43 & 1.965 & 186.5 & 50.4 & Moresco, M. 2015\\
		\hline
		44 & 2.3 & 224.0 & 8.0 & Busca,  N. G. et al. 2013\\
		\hline
		45 & 2.34 & 222 & 7 & Delubac , T. et al. 2015\\
		\hline
		46 & 2.36 & 226 & 8 & Font-Ribera, A., et al. 2014\\
		\hline
	\end{tabular}
\end{minipage}  
\begin{center}
\newpage
~~~~~~~~~~~~~~Table-II : Hubble parameter $H(z)$ with redshift and errors $\sigma_H$ from BAO method~~~~~~~~~~~~~~~~~~~~
\end{center}
\begin{minipage}{.5\linewidth}
	\centering
	\begin{tabular}{ ||>{\centering\arraybackslash}m{0.4cm}|>{\centering\arraybackslash}m{1cm}|>{\centering\arraybackslash}m{1cm}|>{\centering\arraybackslash}m{1cm}|>{\centering\arraybackslash}m{3.15cm}|| }		\hline
		Sl No. & z & $H(z)$ & $\sigma(z)$ & Ref. No. \\
		\hline
		1 & 0.24 & 79.69 & 2.99 & Gazta$\tilde{n}aga$,  E. et al. 2009\\
		\hline
		2 & 0.30 & 81.7 & 6.22 & Oka A. et al. 2014\\
		\hline
		3 & 0.31 & 78.18 & 4.74 & Wang Y. et al. 2017\\
		\hline
		4 & 0.34 & 83.8 & 3.66 & Gazta$\tilde{n}aga$,  E. et al. 2009\\
		\hline
		5 & 0.35 & 82.7 & 9.1 & Chuang, C.H. \& Wang, Y. 2013\\
		\hline
		6 & 0.36 & 79.94 & 3.38 & Wang Y. et al. 2017\\
		\hline
		7 & 0.38 & 81.5 & 1.9 & Alam, S., et al. 2016\\
		\hline
		8 & 0.40 & 82.04 & 2.03 & Wang Y. et al. 2017\\
		\hline
		9 & 0.43 & 86.45 & 3.97 & Gazta$\tilde{n}aga$,  E. et al. 2009\\
		\hline
		10 &  0.44 & 82.6 & 7.8 & Blake, C. et al. 2012\\
		\hline
		11 & 0.44 & 84.81 & 1.83 & Wang Y. et al. 2017\\
		\hline
		12 & 0.48 & 87.79 & 2.03 & Wang Y. et al. 2017\\
		\hline
		13 & 0.51 & 90.4 & 1.9 & Alam, S., et al. 2016\\
		\hline
	\end{tabular}
\end{minipage}
\begin{minipage}{.5\linewidth}
	\centering
	\begin{tabular}{ ||>{\centering\arraybackslash}m{0.4cm}|>{\centering\arraybackslash}m{1cm}|>{\centering\arraybackslash}m{1cm}|>{\centering\arraybackslash}m{1cm}|>{\centering\arraybackslash}m{3.3cm}|| }
 
		\hline
		Sl No. & z & $H(z)$ & $\sigma(z)$ & Ref. No. \\ [0.5ex]
		\hline
		14 & 0.52 & 94.35 & 2.64 & Wang Y. et al. 2017 \\
		\hline 
		15 & 0.56 & 93.34 & 2.3 & Wang Y. et al. 2017 \\
		\hline
		16 & 0.57 & 87.6 & 7.8 & Chuang C-H. et al. 2013 \\
		\hline
		17 & 0.57 & 96.8 & 3.4 & Anderson, L. et al. 2014 \\
		\hline
		18 & 0.59 & 98.48 & 3.18 & Wang Y. et al. 2017 \\
		\hline
		19 & 0.60 & 87.9 & 6.1 & Blake, C. et al. 2012\\
		\hline
		20 & 0.61 & 97.3 & 2.1 & Alam, S., et al. 2016\\
		\hline
		21 & 0.64 & 98.82 & 2.98 & Wang Y. et al. 2017\\
		\hline
		22 & 0.73 & 97.3 & 7.0 & Blake, C. et al. 2012\\
		\hline
		23 & 2.30 & 224 & 8.6 & Busca,  N. G. et al. 2013 \\
		\hline
		24 & 2.33 & 224 & 8 7 & Bautista, J. E. et al. 2017\\
		\hline
		25 & 2.34 & 222 & 8.5 & Delubac , T. et al. 2015 \\
		\hline
		26 & 2.36 & 226 & 9.3 & Font-Ribera, A., et al. 2014\\
		\hline
    \end{tabular}
\end{minipage}	

\begin{figure}[ht]
	\begin{center}
$~~~~~~~~~~~~~~~~~Fig. 1~~~~~~~~~~~~~~~~~~~~~$\\
\includegraphics[height=2in,width=3in]{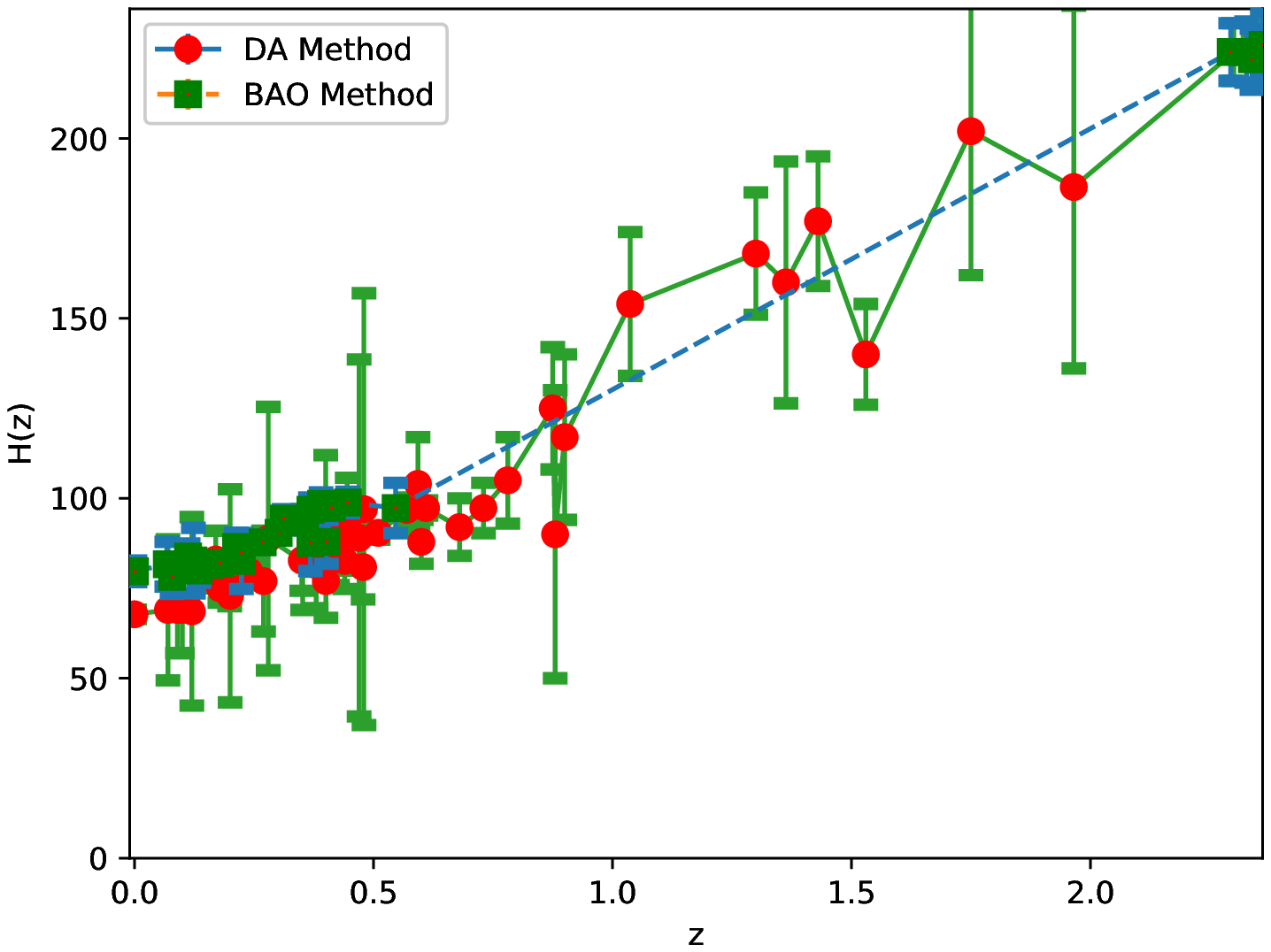}\\

Fig. 1 : $H(z)-z$ graph using the data set from table I and II. Red circle and green rectangle symbol indicate data points of DA method and BAO method respectively.  
\end{center}
\end{figure}
Using the above data for both methods, we will plot $H(z)-z$ graphs in figure 1.

We can see the values of $H(z)$ corresponding to the values of redshift $z$ are semi increasing. So the resultant graph increases with respect to $z$. For lower redshift, BAO method estimates the higher values of $H(z)$ than DA method. Again, for higher redshift, DA method determines the higher values of $H(z)$ as compared to BAO method. Now we will constrain our model's parameters given in equation \eqref{newEoSPB} with the help of the data of table I and II.
\begin{figure}[ht]
    \begin{center}
        $~~~~~~~~~~~~~~~~~~~~~~~Fig.2(a)~~~~~~~~~~~~~~~~~~~~~~~~~~~~~~~~~~~~~~~~~~~~~~~~~~Fig.2(b)~~~~~~~~~~~~~~~~$\\
        \includegraphics[height=2.5in,width=2.8in]{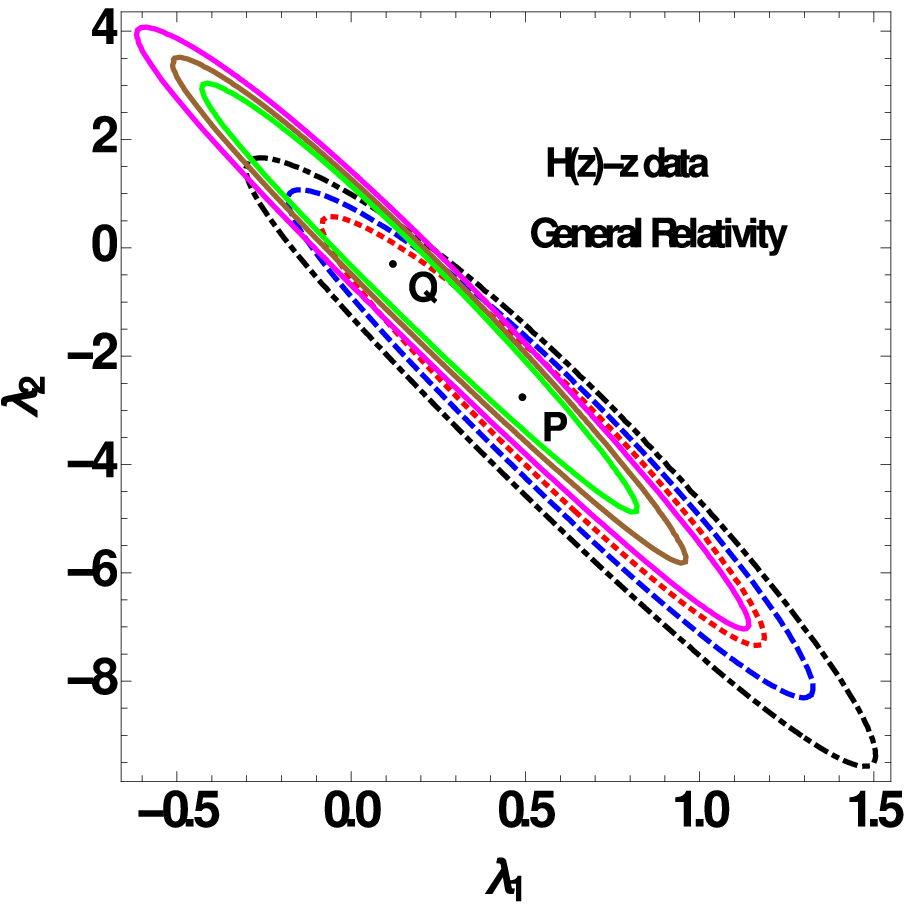} \includegraphics[height=2.5in,width=2.8in]{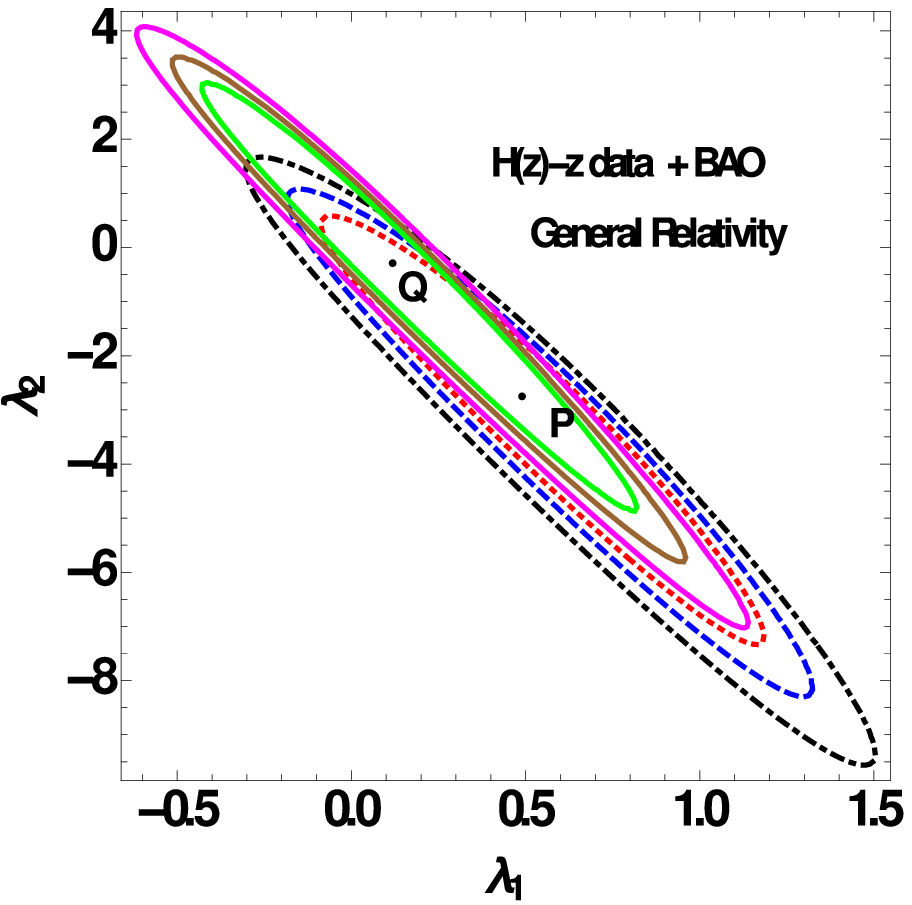}\\
        Fig.2(a)-(b) : $1\sigma$, $~2\sigma$,$~3\sigma$ confidence contours for $H(z)-z$ data set and $H(z)-z$ + BAO data set
        \end{center}
\end{figure}
\begin{wrapfigure}{L}{0.53\textwidth}
    \begin{center}
        $~~~~~~~Fig.2(c)~~~~~~~~~~$\\
        \includegraphics[height=2.5in,width=2.8in]{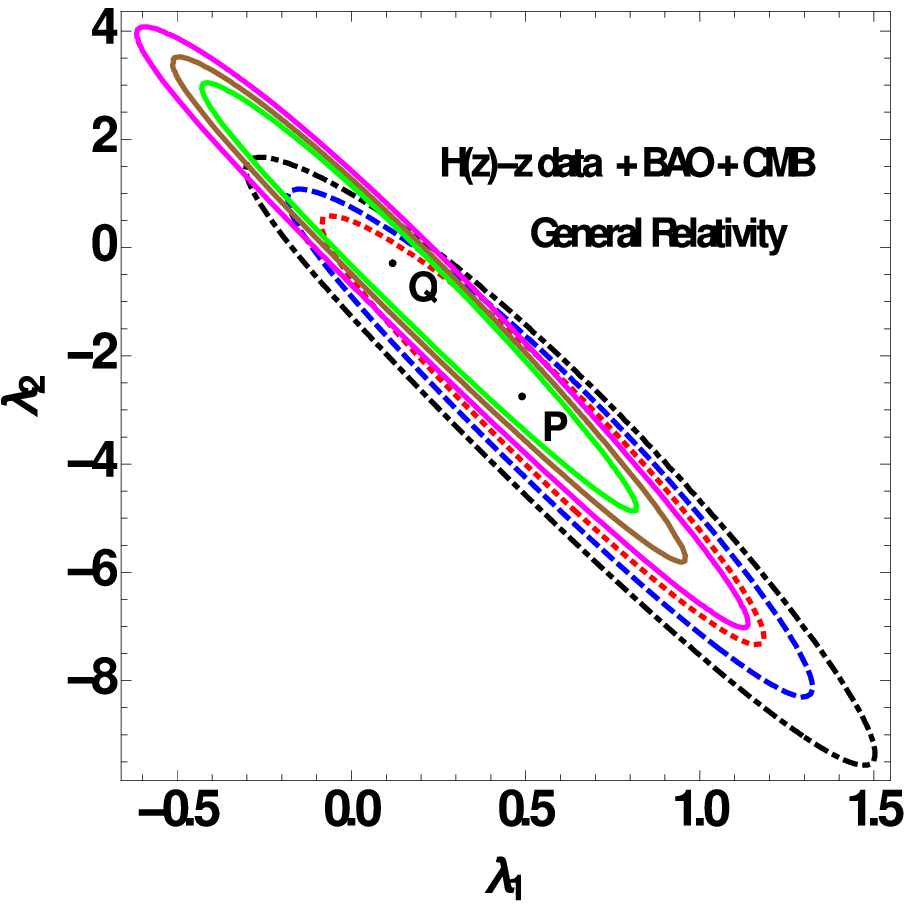}\\
        Fig.2(c) : $1\sigma$, $~2\sigma$,$~3\sigma$ confidence contours for $H(z)-z$ data set + BAO + CMB\
    \end{center}
\end{wrapfigure}

In fig.2(a), we have plotted the best fit values for $\lambda_1$ and $\lambda_2$ along with their corresponding $1\sigma$, $~2\sigma$,$~3\sigma$ confidence contours using the data sets from table I and table II respectively keeping $k_1=k_2=0.1$. Figure $2(b)$ uses BAO with the data sets to constrain $\lambda_1$ and $\lambda_2$. BAO and CMB along with the data sets are given figure 2(c). First we will analyse fig 2(a). For DA method, the point of best fit is denoted by $``P"$ and $1\sigma$, $~2\sigma$,$~3\sigma$ confidence regions are drawn in red, blue and black dashed lines respectively. For BAO method, $``Q"$ is the best fit  and the regions coloured as green, brown and red solid lines for $1\sigma$, $~2\sigma$,$~3\sigma$ confidence contours respectively.

For both the data sets the confidence regions are of more or less elliptic structure and the semi major axis for the regions have the slope more than a right angle. However, slope of confidence contours for DA method is higher than that of BAO method. Tendency wise, high $\lambda_1$ is supported with low $\lambda_2$ to stay in to particular confidence region whereas low value of $\lambda_1$ is accompanied with high $\lambda_2$. This signifies that the DE model given in equation \eqref{EoSPB} allows either of the second or third term to dominate. At the best fits, if we take $z = 0$, then $\omega(z) = -0.553388$ for DA method and $\omega(z) =-0.891208$ for BAO method. The spans of different regions for both cases are enlisted in table - III.

Comparison shows that the length of major axis for confidence contours of BAO data (table I) is less than that for DA data (table II). So lesser amount of region is enclosed as $1\sigma$ confidence if we consider BAO method. The maximum portion of $1\sigma$ region of BAO is common to that of DA. So BAO method has more tendency to constrain the parameters.

In reference of Eisenstein, D. J. et al. (2005), BAO peak parameter is suggested. BAO signal at a scale of $\sim$ $100MPc$ was detected. Using the BAO peak joint analysis, here we will look into the values of  $\lambda_1$ and $\lambda_2$ for our redshift parametrization model. We will analyse our model in the range of $0  <  z <  z_b$, where $z_b = 2.36$ which is called typical redshift (Doran, M. et al 2007) while we use SDSS data samples. BAO peak parameter is defined as

$$\mathscr{A}_{\mathscr{BAO}} = \sqrt{\Omega_m}\bigg(\frac{\int_0^{z_b} E(z)^{-1}dz}{\sqrt{E(z_b)} z_b}\bigg)^{\frac{2}{3}}~~~~~~~~~~.$$   
For flat FLRW model, $\mathscr{A}_{\mathscr{BAO}}~ = ~0.469 \pm 0.017$. Hence, for our analysis, the $\chi^2_{BAO}$ can be written as, 
$$\chi^2_{\mathscr{BAO}} = \frac{(\mathscr{A}_{\mathscr{BAO}} - 0.469)^2}{0.017^2}~~~~~.$$ 

We have plotted the figure 2(b) for $H(z)-z$ data sets along with BAO constraint and the addition of it does not change the basic nature of the confidence curves. The best fits and regions are enlisted are in table IV. In this case, $\omega(z)|_{z=0}$ is $-0.554712$ for DA method and $ -0.892417$ for BAO method.

CMB power spectrum's shift parameter peak is given by  (Efstathiou, G. \& Bond J. R. 1999; Elgaroy, O. \& Multamaki, T. 2007)
$${\mathscr{R}} = \sqrt{\Omega_m} \int_0^{z_c} \frac{dz'}{E(z')}~~~~~~~~~~,$$
where $z_c$ describes the value of the particular $z$ at the last scattering surface. WMAP predicts the value of ${\mathscr{R}} = 1.726 \pm 0.018$ at $z_c = 1091.3$. The $\chi^2_{\mathscr{CMB}}$ function (for CMB measurement) is defined as

$$\chi^2_{\mathscr{CMB}} = \frac{({\mathscr{R}}- 1.726)^2}{0.018^2}~~~~~~~~~~.$$

We have imposed this method and have plotted the confidence contours in fig 2(c).

The best fits and the spans of confidence contour curves are mentioned in the table V. Secondly, it is noticeable that the best fits are placed in the fourth quadrant. In this case at present epoch, i.e., at $z = 0$, $\omega(z) = -1$ for DA method and $-0.893717$ for BAO method.

As $\omega(z)|_{z=0} = -1$ after using BAO and CMB in our model with $H(z)-z$ data from DA method, it is significant that our cosmological model with BAO and CMB is the best measurement to support our result with $\Lambda$CDM model.

We will vary $k_1$ and $k_2$ to draw $\lambda_1$-$\lambda_2$ confidence contours in figures 2(d)-(g).

Data set obtained by DA method is used to plot fig 2(d) (varying $k_1$, $k_2$ is fixed at 0.1) and fig. 2(e) (varying $k_2$, $k_1$ is fixed at 0.1). If $k_2$ is fixed and $k_1$ is increased it is observed that the regions of the confidence contours increase along with a rotation in the semi major axis with positive angles. The corner of low $\lambda_1$ high $\lambda_2$ is fixed as the center of rotation. This signifies high $\lambda_1$ and low $\lambda_2$ values are permissible if $k_1$ is increased. If $k_2$ is increased and $k_1$ is kept to be fixed, angle of rotation of the semi major axis of the confidence regions. So increment in $k_2$ prefer low $\lambda_1$ with low $\lambda_2$ to stay in the same confidence. In this case, the center of rotation also stays in the low $\lambda_1$ - high $\lambda_2$ region.

Confidence contours obtained from BAO method's data set are plotted in fig 2(f) ($k_2=0.1$, varying $k_1$) and fig 2(g) ($k_2$ varying, $k_1=0.1$). Basic natures do match with 2(d) and 2(e) respectively. Only the difference is the center of rotation for major axis is near by best fit values. So if $k_1$ is increased BAO data support high $\lambda_1$ - low $\lambda_2$ regions along with low $\lambda_1$ - high $\lambda_2$ regions. On the otherhand increment of $k_2$ supports low $\lambda_1$ - low $\lambda_2$ or high $\lambda_1$ - high $\lambda_2$ regions. For both the data sets it is observed that if $k_1$ and $k_2$ are low, constraining is more stronger.

\vspace{.1 in}

\begin{center}
	Table III :The best fit values of $\lambda_1$, $\lambda_2$, $\chi^2$ and corresponding region of $1\sigma$, $2\sigma$ and $3\sigma$ for both DA and BAO method using $H(z)-z$ data set with $k_1=k_2=0.1$
	\begin{tabular}{| >{\centering\arraybackslash}m{2.2cm} | >{\centering\arraybackslash}m{1.2cm}| >{\centering\arraybackslash}m{2.5cm}| >{\centering\arraybackslash}m{2.5cm}| >{\centering\arraybackslash}m{2.5cm} | >{\centering\arraybackslash}m{2.5cm} |}
		\hline
		Tools & stat. info & \multicolumn{4}{c|}{Region of the contours}  \\ 
		\hline
		\multirow{6}{*}{$H(z)-z$ data} & \multirow{3}{*}{Best fits}  & \multicolumn{2}{c|}{DA Method} & \multicolumn{2}{c|}{BAO Method}    \\ [0.5ex]
		\cline{3-6}
		&  & \multicolumn{2}{c|}{$\chi^2 = 44.3734~,~\omega(z)\big|_{z=0}=-0.553388$}  & \multicolumn{2}{c|}{$\chi^2 = 20.0144~,~\omega(z)\big|_{z=0}=-0.891208$}  \\
		\cline{3-6}
		& & $\lambda_1 = 0.491273~~(V_{11})$ & $\lambda_2 = -2.75766~~(V_{12})$ &  $\lambda_1 = 0.119671~~(V_{21})$ & $\lambda_2 = -0.297328~~(V_{22})$    \\
		\cline{2-6}
		&$1\sigma$ & ${V_{11}}^{+0.704727}_{-0.587388}$ & ${V_{12}}^{+3.38546}_{-4.64334}$ & ${V_{21}}^{+0.712429}_{-0.566171}$  & ${V_{22}}^{+3.392328}_{-4.689672}$ \\
		\cline{2-6}
		&$2\sigma$ & ${V_{11}}^{+0.846727}_{-0.684273}$  & ${V_{12}}^{+3.87666}_{-5.62634}$ & ${V_{21}}^{+0.853029}_{-0.646471}$ & ${V_{22}}^{+3.867328}_{-5.600672}$ \\
		\cline{2-6}
		&$3\sigma$ & ${V_{11}}^{+1.021727}_{-0.800673}$& ${V_{12}}^{+4.49166}_{-6.85434}$ & ${V_{21}}^{+1.033329}_{-0.746871}$ & ${V_{22}}^{+4.461328}_{-6.789672}$ \\
		\hline
	\end{tabular}
\end{center}
\begin{center}
	Table IV :The best fit values of $\lambda_1$, $\lambda_2$, $\chi^2$ and corresponding region of $1\sigma$, $2\sigma$ and $3\sigma$ for both DA and BAO method using $H(z)-z$ + BAO data set $k_1=k_2=0.1$
	\begin{tabular}{| >{\centering\arraybackslash}m{2.2cm} | >{\centering\arraybackslash}m{1.2cm}| >{\centering\arraybackslash}m{2.5cm}| >{\centering\arraybackslash}m{2.5cm}| >{\centering\arraybackslash}m{2.5cm} | >{\centering\arraybackslash}m{2.5cm} |}
		\hline
		Tools & stat. info & \multicolumn{4}{c|}{Region of the contours}  \\ 
		\hline
		\multirow{6}{*}{\makecell{$H(z)-z$ data \\ + BAO}} & \multirow{3}{*}{Best fits}  & \multicolumn{2}{c|}{DA Method} & \multicolumn{2}{c|}{BAO Method}    \\ [0.5ex]
		\cline{3-6}
		&  & \multicolumn{2}{c|}{$\chi^2 = 804.712~,~\omega(z)\big|_{z=0} = -0.554712$}  & \multicolumn{2}{c|}{$\chi^2 = 780.327~,~\omega(z)\big|_{z=0} = -0.892417$}  \\
		\cline{3-6}
		& & $\lambda_1 = 0.489817 ~(V_{11})$ & $\lambda_2 = -2.75048 ~(V_{12}) $ &  $\lambda_1 = 0.118341 ~(V_{21})$ & $\lambda_2 = -0.289766~ (V_{22})$    \\
		\cline{2-6}
		&$1\sigma$ & ${V_{11}}^{+7.06183}_{-0.590517}$ & ${V_{12}}^{+3.373787}_{-4.614513}$ & ${V_{21}}^{+0.714359}_{-0.564441}$  & ${V_{22}}^{+3.420766}_{-4.667234}$ \\
		\cline{2-6}
		&$2\sigma$ & ${V_{11}}^{+0.846183}_{-0.685317}$  & ${V_{12}}^{+3.914487}_{-5.618513}$ & ${V_{21}}^{+0.852759}_{-0.650141}$ & ${V_{22}}^{+3.900766}_{-5.585234}$ \\
		\cline{2-6}
		&$3\sigma$ & ${V_{11}}^{+1.025183}_{-0.794917}$& ${V_{12}}^{+4.454487}_{-6.853513}$ & ${V_{21}}^{+1.037659}_{-0.742441}$ & ${V_{22}}^{+4.469766}_{-6.809234}$ \\
		\hline
	\end{tabular}
\end{center}
\newpage
\begin{center}
	Table V :The best fit values of $\lambda_1$, $\lambda_2$, $\chi^2$ and corresponding region of $1\sigma$, $2\sigma$ and $3\sigma$ for both DA and BAO method using $H(z)-z$ + BAO + CMB data set $k_1=k_2=0.1$
	\begin{tabular}{| >{\centering\arraybackslash}m{2.2cm} | >{\centering\arraybackslash}m{1.2cm}| >{\centering\arraybackslash}m{2.5cm}| >{\centering\arraybackslash}m{2.5cm}| >{\centering\arraybackslash}m{2.5cm} | >{\centering\arraybackslash}m{2.5cm} |}
		\hline
		Tools & stat. info & \multicolumn{4}{c|}{Region of the contours}  \\ 
		\hline
		\multirow{6}{*}{\makecell{$H(z)-z$ data \\ + BAO \\ + CMB}} & \multirow{3}{*}{Best fits}  & \multicolumn{2}{c|}{DA Method} & \multicolumn{2}{c|}{BAO Method}    \\ [0.5ex]
		\cline{3-6}
		&  & \multicolumn{2}{c|}{$\chi^2 = 9999.38~,~\omega(z)\big|_{z=0} = -1$}  & \multicolumn{2}{c|}{$\chi^2 = 9974.99~,~\omega(z)\big|_{z=0} = -0.893717$}  \\
		\cline{3-6}
		& & $\lambda_1 = 0.489817~ (V_{11})$ & $\lambda_2 = -2.75048 ~(V_{12}) $ &  $\lambda_1 = 0.118341 ~(V_{21})$ & $\lambda_2 = -0.289766~ (V_{22})$    \\
		\cline{2-6}
		&$1\sigma$ & ${V_{11}}^{+0.700183}_{-0.582677}$ & ${V_{12}}^{+3.38758}_{-4.63252}$ & ${V_{21}}^{+0.720459}_{-0.564841}$  & ${V_{22}}^{+3.344766}_{-4.658234}$ \\
		\cline{2-6}
		&$2\sigma$ & ${V_{11}}^{+0.846183}_{-0.683817}$  & ${V_{12}}^{+3.86148}_{-5.61952}$ & ${V_{21}}^{+0.861059}_{-0.645141}$ & ${V_{22}}^{+3.899766}_{-5.608234}$ \\
		\cline{2-6}
		&$3\sigma$ & ${V_{11}}^{+1.016183}_{-0.803917}$& ${V_{12}}^{+4.41448}_{-6.88352}$ & ${V_{21}}^{+1.041659}_{-0.745541}$ & ${V_{22}}^{+4.414766}_{-6.797234}$ \\
		\hline
	\end{tabular}
\end{center}
\begin{center}
	Table VI :The best fit values of $\lambda_1$, $\lambda_2$, $\chi^2$ from $1\sigma$, $2\sigma$ and $3\sigma$ confidence contours for both DA and BAO method using $H(z)-z$ data set and different $k_1$ and $k_2$ values.\\
	\begin{tabular}{|| >{\centering\arraybackslash}m{1.5cm}  >{\centering\arraybackslash}m{1.5cm} >{\centering\arraybackslash}m{1.5cm} >{\centering\arraybackslash}m{1.5cm} ||>{\centering\arraybackslash}m{1.5cm}  >{\centering\arraybackslash}m{1.5cm}  >{\centering\arraybackslash}m{1.5cm}  >{\centering\arraybackslash}m{1.5cm} ||}
	\hline
	\multicolumn{8}{||c||}{DA method}  \\
	\hline
	\multicolumn{4}{||c||}{$k_2~=~0.1$} & \multicolumn{4}{|c||}{$k_1~=~0.1$}  \\
	\hline
	$k_1$ & $\lambda_1$ & $\lambda_2$ & $\chi^2$ & $k_2$ & $\lambda_1$ & $\lambda_2$ & 			$\chi^2$  \\ 
	\hline
	$-0.99$ & $0.0217069$ & $-1.32615$ & $46.0804$ & $-0.99$ & $-0.473434$ & $0.0607605$ & $47.9418$ \\
	
	$-0.5$ & $0.229342$ & $-2.36384$ & $44.5134$ & $-0.5$ & $0.67951$ & $-1.28682$ & $46.7378$ \\ 
	
	$0.1$ & $0.491273$ & $-2.75766$ & $44.3734$ & $0.1$ & $0.491273$ & $-2.75766$ & $44.3734$ \\
	
	$1$ & $0.900393$ & $-3.03312$ & $44.4522$ & $1$ & $0.414204$ & $-5.98958$ & $42.9764$ \\
	
	$2$ & $1.36386$ & $-3.18998$ & $44.577$ & $2$ & $0.379361$ & $-10.9337$ & $42.2397$\\
	\hline
	\multicolumn{8}{||c||}{BAO method}  \\
	\hline
	\multicolumn{4}{||c||}{$k_2~=~0.1$} & \multicolumn{4}{|c||}{$k_1~=~0.1$}  \\
	\hline
	$k_1$ & $\lambda_1$ & $\lambda_2$ & $\chi^2$ & $k_2$ & $\lambda_1$ & $\lambda_2$ & 			$\chi^2$  \\ 
	\hline
	$-0.99$ & $0.00478714$ & $0.103182$ & $20.2056$ & $-0.99$ & $0.113188$ & $-0.00458441$ & 20.0675 \\
	
	$-0.5$ & $0.0548766$ & $-0.188829$ & $20.0389$ & $-0.5$ & $0.0551597$ & $0.0382559$ & $20.0798$ \\ 
	
	$0.1$ & $0.119671$ & $-0.297328$ & $20.0144$ & $0.1$ & $0.119671$ & $-0.297328$ & $20.0144$ \\
	
	$1$ & $0.216863$ & $-0.354743$ & $20.0289$ & $1$ & $0.138063$ & $-1.11864$ & $19.8405$ \\
	
	$2$ & $0.322417$ & $-0.376157$ & $20.0515$ & $2$ & $0.143973$ & $-2.49855$ & $19.694$\\
	\hline
	\end{tabular}
\end{center}	
\begin{figure}[ht]
    \begin{center}
$~~~~~~~~~~~~~~~~~~~~~~~Fig.2(d)~~~~~~~~~~~~~~~~~~~~~~~~~~~~~~~~~~~~~~~~~~~~~~~~~~Fig.2(e)~~~~~~~~~~~~~~~~$
\includegraphics[height=2.5in,width=2.8in]{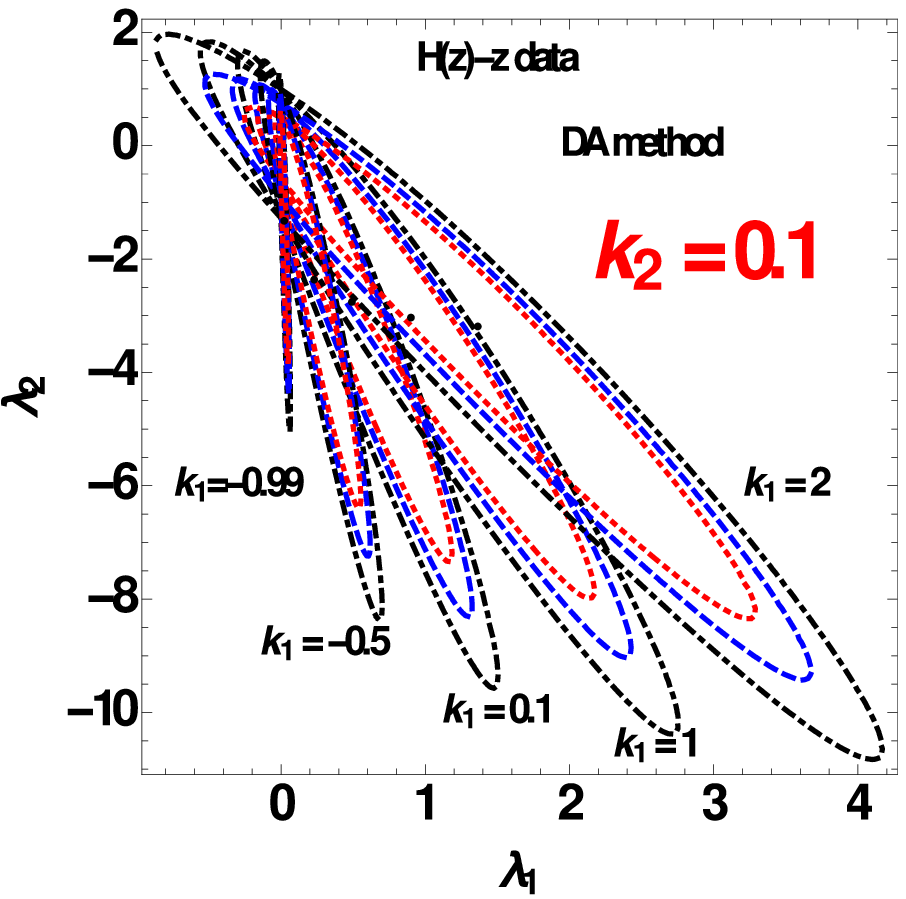} 
\includegraphics[height=2.5in,width=2.8in]{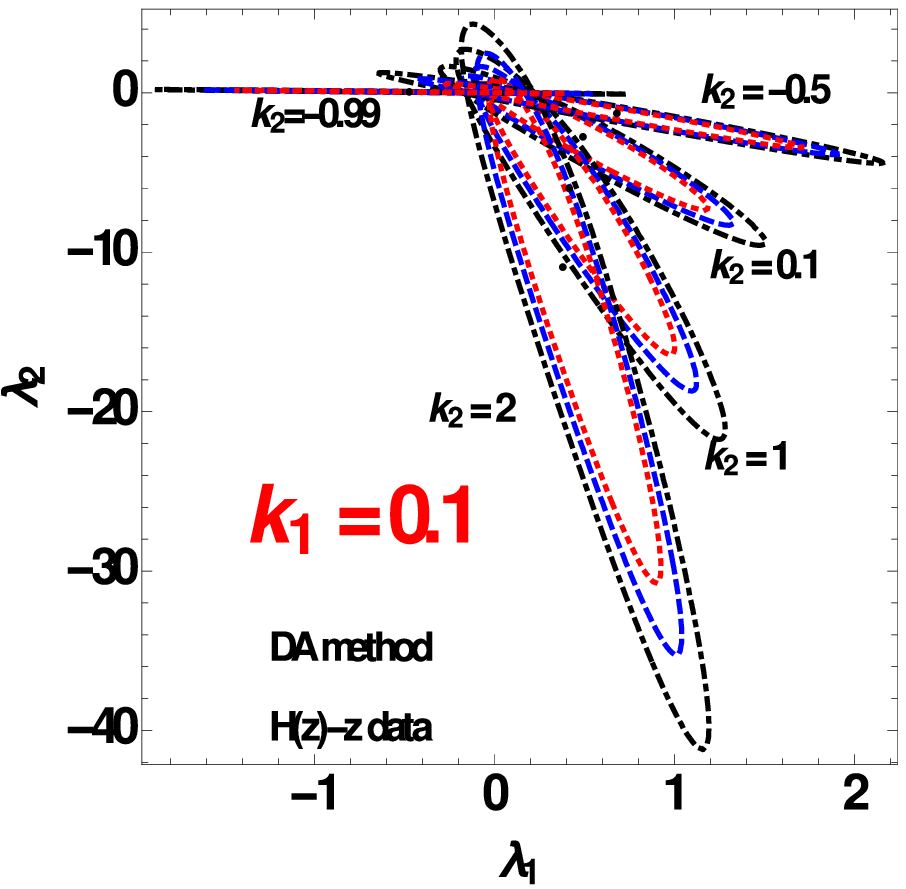} \\
$~~~~~~~~~~~~~~~~~~~~~~~Fig.2(f)~~~~~~~~~~~~~~~~~~~~~~~~~~~~~~~~~~~~~~~~~~~~~~~~~~Fig.2(g)~~~~~~~~~~~~~~~~$
\includegraphics[height=2.5in,width=2.8in]{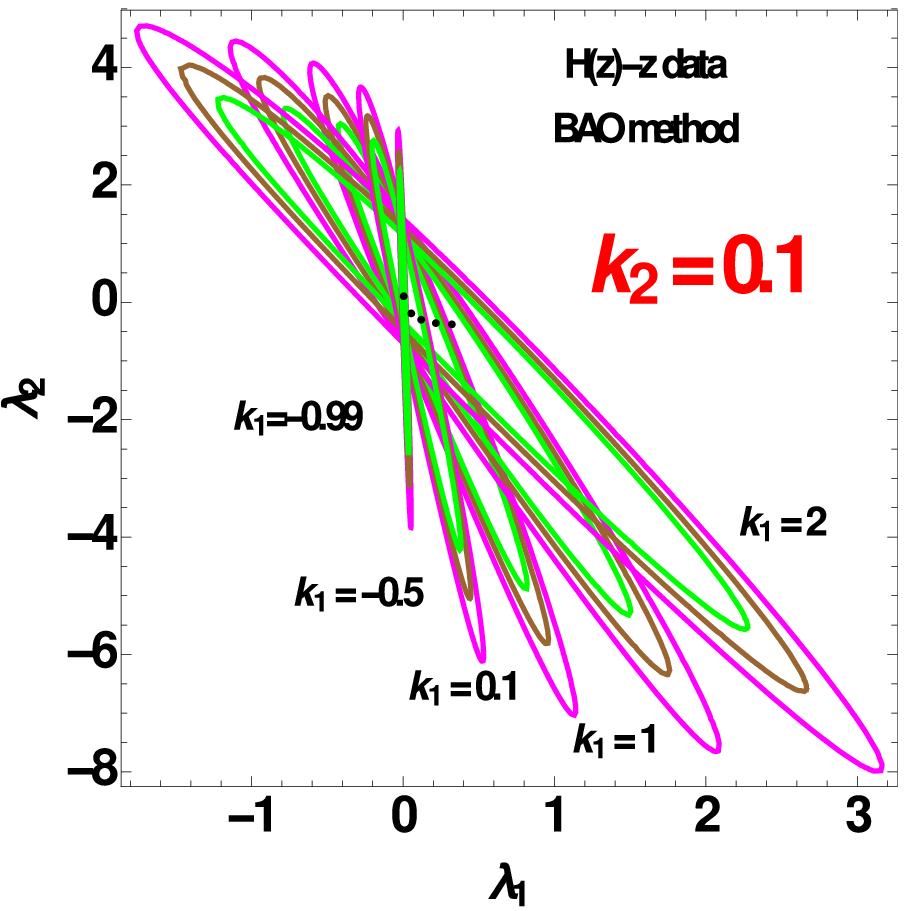}
\includegraphics[height=2.5in,width=2.8in]{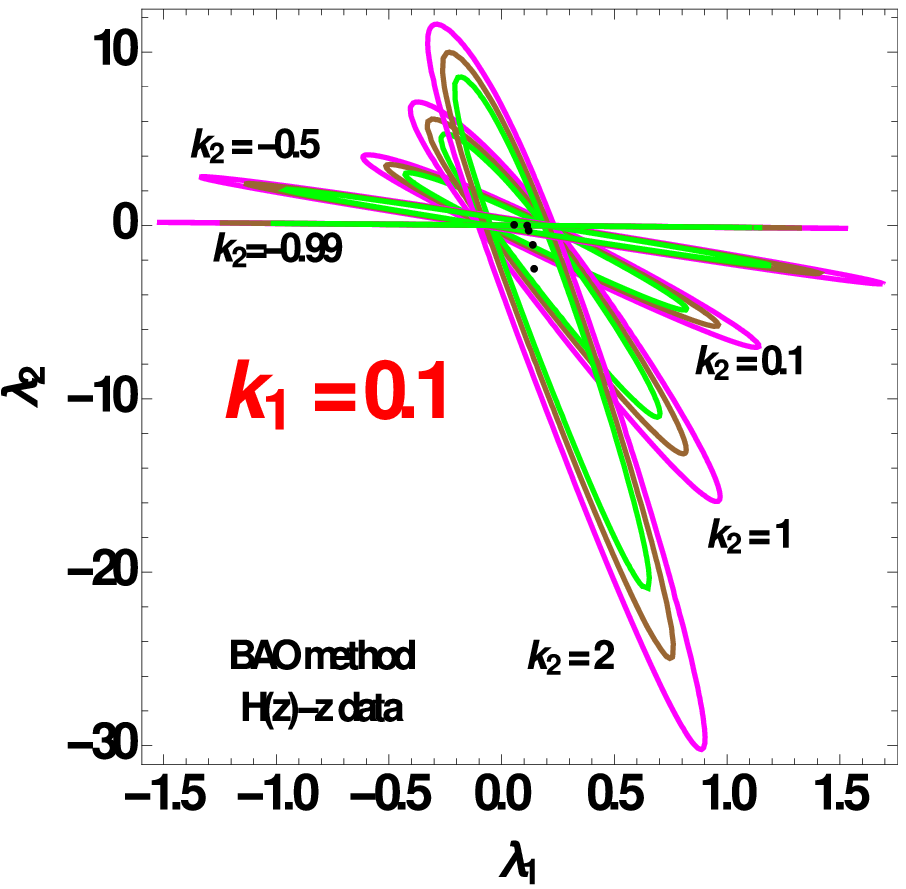}\\ 
Fig.2(d)-(g) : $1\sigma$, $~2\sigma$,$~3\sigma$ confidence contours in $\lambda_1$-$\lambda_2$ plane for $H(z)-z$ data set obtained by DA method (fig.2(d) and fig.2(e)) and BAO method (fig.2(f) and fig.2(g)) for different $k_1$ (say fig.2(d) and fig.2(e)) and $k_2$ (say fig.2(f) and fig.2(g)) values.
        \end{center}
\end{figure}

\section{Studies of Different Cosmological Parameters :}
The deceleration parameter $q$ is defined as
\begin{equation}\label{formula_deceleration_parameter}
q = -\frac{\ddot{a}}{aH^2} = -\bigg(1 + \frac{\dot{H}}{H^2}\bigg)~~~~~~~,
\end{equation}
where $\dot{H} = \frac{dH}{dt} = aH\frac{dH}{da}$.

From equations (\ref{Obtained_Hubbles_parameter}) and (\ref{formula_deceleration_parameter}), the expression for $q$ in terms of scale factor $a$ can be written as,
\begin{equation}\label{Obtained_deceleration_parameter_a}
 q(a) = -1 + \frac{2\Omega_{rad0}a^{-4} + \frac{3}{2}\Omega_{dm0}a^{-3}- \frac{3a}{2}\beta~\Omega_{c\phi 0}~\frac{\left(1+ak_2\right)^{\frac{3\lambda_2}{k_2^2}}}{\left(1+ak_1\right)^{\frac{3\lambda_1}{k_1}}}\bigg\{\frac{\lambda_2}{1+ak_2} - \frac{\lambda_2~(1+k_2)}{(1+ak_2)^{2}}  - \frac{\lambda_1}{1+ak_1} \bigg\} exp\bigg\{\frac{3\lambda_2(1+k_2)}{k_2^2(1+ak_2)}\bigg\}}{\Omega_{rad0}a^{-4} + \Omega_{dm0}a^{-3} + \beta~\Omega_{c\phi 0}\frac{\left(1+ak_2\right)^{\frac{3\lambda_2}{k_2^2}}}{\left(1+ak_1\right)^{\frac{3\lambda_1}{k_1}}}exp\bigg\{\frac{3\lambda_2(1+k_2)}{k_2^2(1+ak_2)}  \bigg\}}~~.
\end{equation}
\newpage
Now, the equation (\ref{Obtained_deceleration_parameter_a}), in terms of redshift $z$ is
$$ q(z) = -1 + \left[\Omega_{rad0}(1+z)^4 + \Omega_{dm0}(1+z)^3 + \beta\Omega_{c\phi 0}\left(\frac{1+z}{1+k_1+z}\right)^{\frac{3 \lambda_1}{k_1}}\left(\frac{1+k_2+z}{1+z}\right)^{\frac{3 \lambda_2}{k_2^2}} exp\bigg\{\frac{3\lambda_2(1+k_2)(1+z)}{k_2^2(1+k_2+z)} \bigg\}\right]^{-1}
$$
$$~~~~~~~~~~~~~~~~~~~~~\times\left[2\Omega_{rad0}(1+z)^4 + \frac{3}{2}\Omega_{dm0}(1+z)^3- \frac{3\beta}{2}\Omega_{c\phi 0}\bigg\{ ~\frac{\lambda_2}{1+k_2+z} - \frac{\lambda_2~(1+k_2)(1+z)}{(1+k_2+z)^2} - \frac{\lambda_1}{1+k_1+z} \bigg\}\right.
$$
\begin{equation}\label{Obtained_deceleration_parameter_z}
~~~~~~~~~~~~~~~~~~~~~~~~~~~~~~~~~~~~~~~~~~~~~~~~~~~~~~~~~~~~~\left.\times \left(\frac{1+z}{1+k_1+z}\right)^{\frac{3 \lambda_1}{k_1}}\left(\frac{1+k_2+z}{1+z}\right)^{\frac{3 \lambda_2}{k_2^2}} exp\bigg\{\frac{3\lambda_2(1+k_2)(1+z)}{k_2^2(1+k_2+z)} \bigg\}\right]
\end{equation}
To study the situation in every direction, we analyse the density parameters for the matter field ($\Omega_{dm}$) and scalar field ($\Omega_{c\phi}$) as,
\begin{equation}
\Omega_{dm}(z) = \frac{\Omega_{dm0}(1+z)^3}{\Omega_{rad0}(1+z)^4 + \Omega_{dm0}(1+z)^3 + \beta~\Omega_{c\phi 0}~\left(\frac{1+z}{1+k_1+z}\right)^{\frac{3 \lambda_1}{k_1}}\left(\frac{1+k_2+z}{1+z}\right)^{\frac{3 \lambda_2}{k_2^2}} exp
\bigg\{\frac{3\lambda_2(1+k_2)(1+z)}{k_2^2(1+k_2+z)} \bigg\} }~~~~~~~~~~~and
\end{equation}
\begin{equation}
\Omega_{c\phi}(z) = \frac{\beta~\Omega_{c\phi 0} ~ (\frac{1+z}{1+k_1+z})^{\frac{3 \lambda_1}{k_1}}(\frac{1+k_2+z}{1+z})^{\frac{3 \lambda_2}{k_2^2}} exp\bigg\{\frac{3\lambda_2(1+k_2)(1+z)}{k_2^2(1+k_2+z)} \bigg\}}{\Omega_{rad0}(1+z)^4 + \Omega_{dm0}(1+z)^3 + \beta~\Omega_{c\phi 0} ~\left(\frac{1+z}{1+k_1+z}\right)^{\frac{3 \lambda_1}{k_1}}\left(\frac{1+k_2+z}{1+z}\right)^{\frac{3 \lambda_2}{k_2^2}} exp\bigg\{\frac{3\lambda_2(1+k_2)(1+z)}{k_2^2(1+k_2+z)} \bigg\} }~~~~~~~~~~~~~~~~.
\end{equation}
Now adding (\ref{cphidensity3}) and (\ref{scalar_field_equation}), we can obtain,
$$\dot{c\phi}^2 = (1+z)^2 H^2 \bigg(\frac{dc\phi}{dz}\bigg)^2\Rightarrow \frac{dc\phi(z)}{dz} = \pm \sqrt{3}(1+z)^{-1} \bigg[\frac{\lambda_1}{1+k_1+z}+\frac{\lambda_2 z}{(1+k_2+z)^2}\bigg]^{\frac{1}{2}}~~~~~~~~~~~~~~~~~~~~~~~~~~~~~~~~~~~~~~~$$
\begin{equation}
~~~~~~~~~~~~~~~~~~~~~~~~~~~~~~~~~~~~~~~~~~~~~~~~~~~~~~~~~~~~~~~~~~~~~~\times ~\left[1 + \frac{\Omega_{rad0}(1+z)^4+\Omega_{dm0}(1+z)^3}{\Omega_{c\phi 0}\beta \bigg(\frac{1+z}{1+k_1+z}\bigg)^{\frac{3 \lambda_1}{k_1}} \bigg(\frac{1+k_2+z}{1+z}\bigg)^{\frac{3 \lambda_2}{k^2_2}} exp\bigg\{ \frac{3\lambda_2(1+k_2)(1+z)}{k_2^2(1+k_2+z)} \bigg\}}\right]^{-\frac{1}{2}}~~~~~.
\end{equation}
Again from equations (\ref{cphidensity3}) and (\ref{scalar_field_equation}), we can rewrite the potential in terms of that scalar field as,
\begin{equation}
V(c\phi) = \frac{1}{2}\rho_{c\phi}(1-\omega_{c\phi})~~~~~~~~.
\end{equation}
In terms of $z$, we can express $V(z)$ as
\begin{equation}
V(z) = V_0 \bigg(\frac{1+z}{1+k_1+z}\bigg)^{\frac{3 \lambda_1}{k_1}}\bigg(\frac{1+k_2+z}{1+z}\bigg)^{\frac{3 \lambda_2}{k_2^2}} exp\bigg\{\frac{3\lambda_2(1+k_2)(1+z)}{k_2^2(1+k_2+z)}\bigg\}\bigg\{1-\frac{\lambda_1}{2(1+k_1+z)}-\frac{\lambda_2 z}{2(1+k_2+z)^2}\bigg\}~~~~~~~~~,
\end{equation}
where , $V_0 = 3 H_0^2 \Omega_{c\phi 0}\beta$.

\begin{figure}[ht]
\begin{center}
$~~~~~~~~~~~~~Fig.3(a)~~~~~~~~~~~~~~~~~~~~~~~~~~~~~~~~~~~~~~Fig.3(b)~~~~~~~~~~~~~~~~~~~~~~~~~~~~~~~~~~Fig.3(c)~~~~~~~~~~~~~$\\
\includegraphics[height=2.1in,width=2.3in]{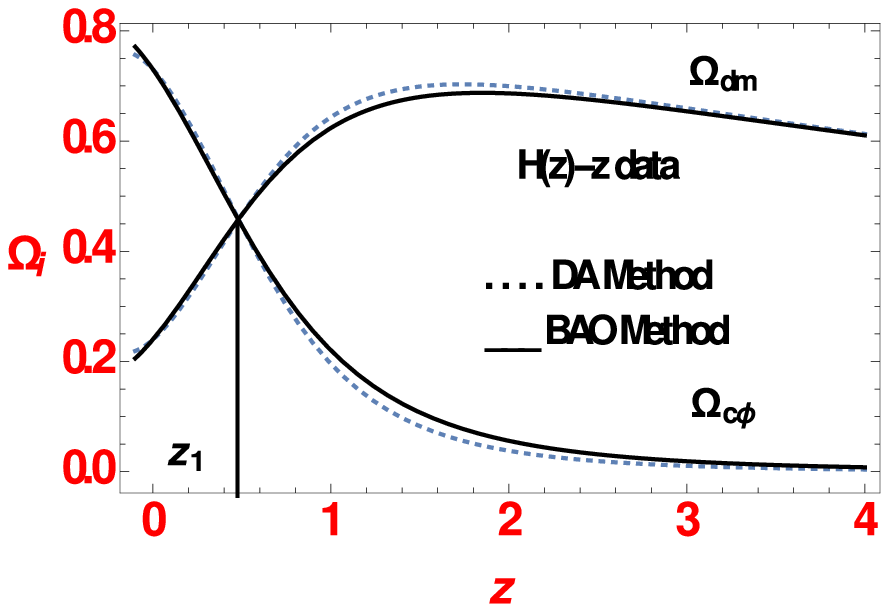}~\includegraphics[height=2.1in,width=2.3in]{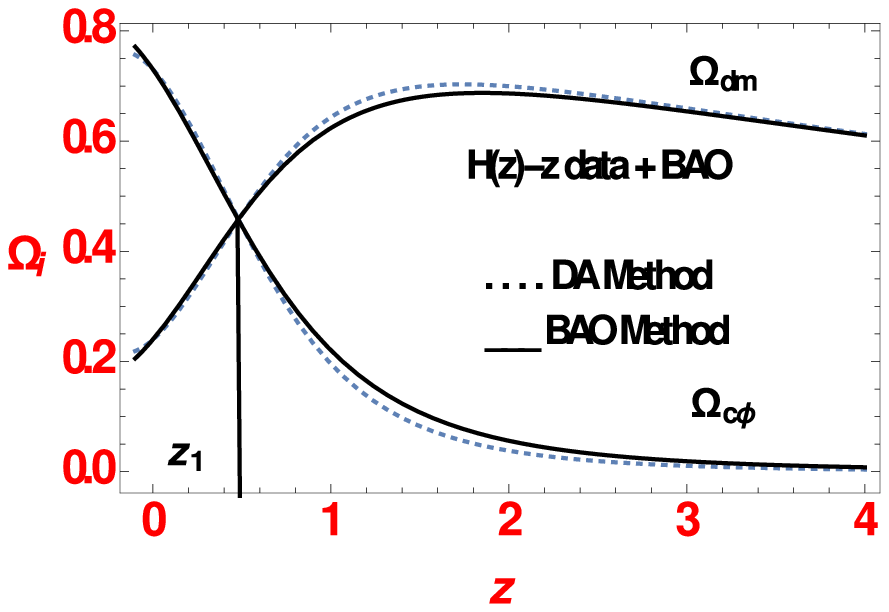}~\includegraphics[height=2.1in,width=2.3in]{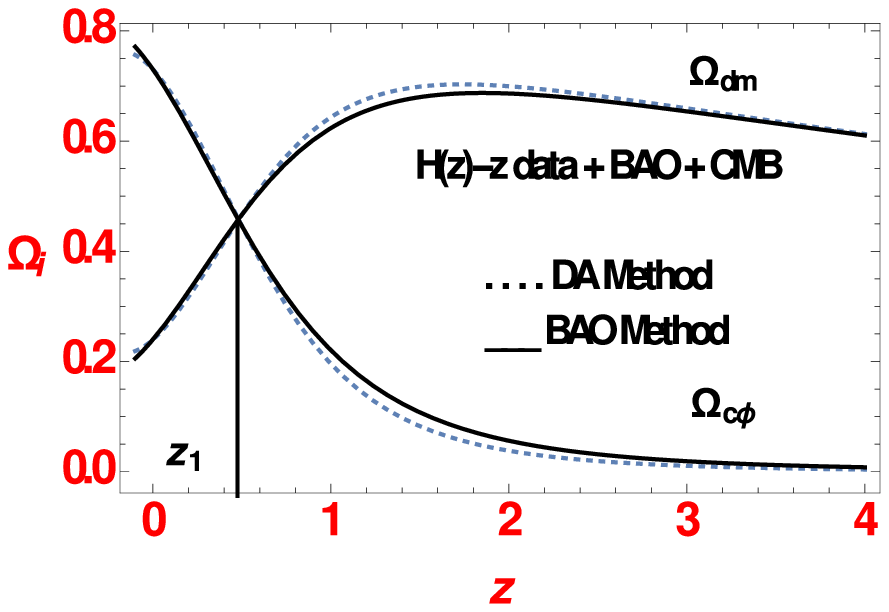}\\
Fig.3(a)-(c) : $\Omega_{dm},~\Omega_{c\phi} $ vs $z$ graphs, where dotted lines show DA method and solid lines show BAO method in the case of $H(z)-z$ data, $H(z)-z$ data with BAO, $H(z)-z$ data + BAO + CMB respectively. \\
\end{center}
\end{figure}
In figure 3(a)-3(c) we have plotted the fractional dimensionless densities $\Omega_{dm}$ for matter and $\Omega_{c\phi}$ for the exotic matter with respect to redshift $z$ for three of the cases, $H(z)-z$ data, $H(z)-z$ data+ BAO, $H(z)-z$ data + BAO + CMB respectively. We observe that the fractional densities are of increasing nature in past with more or less same slope. $\Omega_{dm} > \Omega_{c\phi}$ for high $z$. But as time grows, $\Omega_{c\phi}$ increases and after a certain point, $z=z_1$ (say), $\Omega_{c\phi}$ turns greater than $\Omega_{dm}$. These graphs, to some extents, support the theory that in extreme part, the universe was matter dominated with $\Lambda = 0$. But as time grows, a delayed decay in matter world took place. This converted matter into relativistic hypothetical energy counterpart and finally $\Lambda = -1$ epoch came to exist at the present time. This theory (Turner, M. S. et al 1984; Mathews,  G. J. 2008) even helped a lot to bypass different theoretical discrepancies faced by $\Lambda = -1$ model alone. This is consistent with the results of Pan, S. \& Chakraborty, S. (2013) and Chakraborty, S. et al. (2014). Fig 3(a)-(c) also depict that DA method is more appropriate than BAO to explain the transit from $\Lambda = 0$ to $-1$.
\begin{figure}[ht]
\begin{center}
$~~~~~~~~~~~~~Fig.4(a)~~~~~~~~~~~~~~~~~~~~~~~~~~~~~~~~~~~~~~Fig.4(b)~~~~~~~~~~~~~~~~~~~~~~~~~~~~~~~~~~~~Fig.4(c)~~~~~~~~~~$\\    
\includegraphics[height=2.1in,width=2.3in]{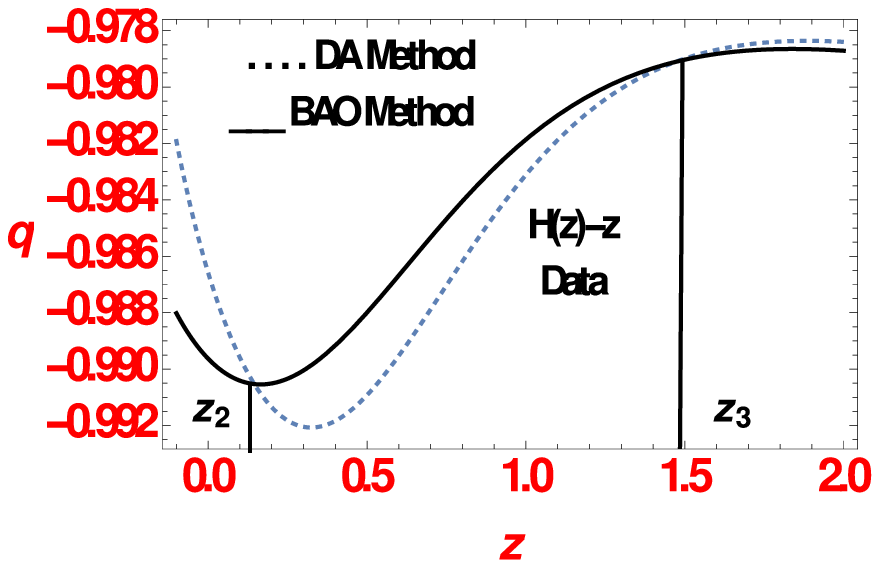}
\includegraphics[height=2.1in,width=2.3in]{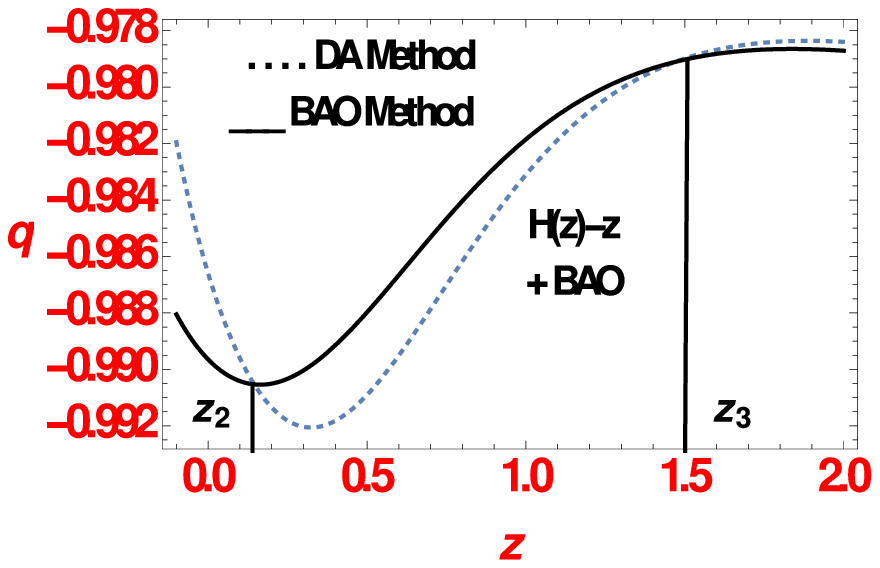}
\includegraphics[height=2.1in,width=2.3in]{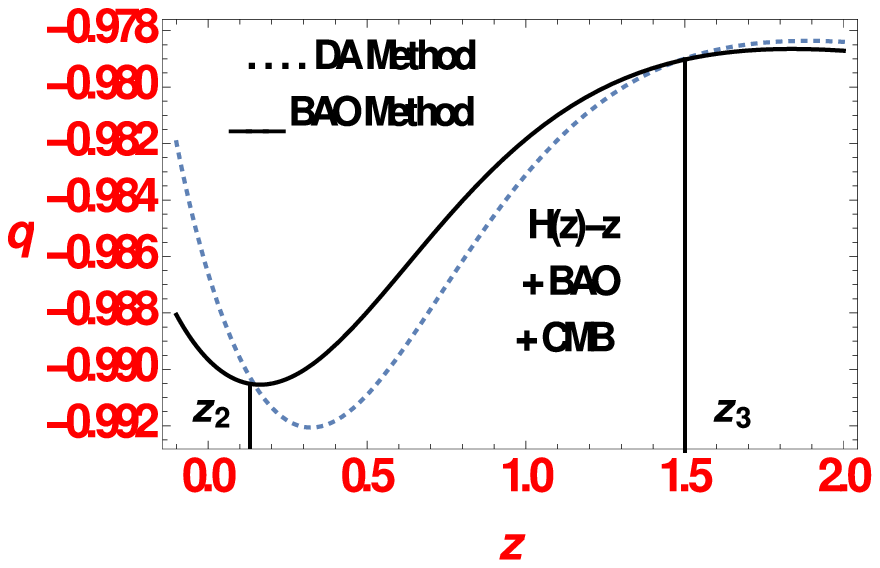}\\
Fig.4(a)-(c) : $q$ vs $z$ graphs, where dotted lines are drawn for DA method and solid lines are drawn for BAO method. From fig 4(a) to 4(c) the graphs are for $H(z)-z$ data, $H(z)-z$ data + BAO and $H(z)-z$ data + BAO + CMB respectively. \\
\end{center}
\end{figure}

We plot deceleration parameter, $q$, as a function of $z$ in fig 4(a)-(c). $q$ is found to be a decreasing function of time. The rate of $q$'s contraction is low at high $z$ and high at low $z$. We do not find any value $z$ where $q$ changes its sign. So a transition from deceleration to acceleration or the opposite is not allowed for our model. We find at least two values of ($z =z_2$ and $z_3$, say) where, as the time grows, i.e., $z$ decreases $q_{BAO}(z) < q_{DA}(z)$ as $z > z_3$, then $q_{BAO}(z) > q_{DA}(z)$ for the interval $z_2 < z < z_3$ and finally $q_{BAO}(z) < q_{DA}(z)$ when $z < z_2$. So $z_2$ and $z_3$ are such two points for which the same deceleration is permitted by both the data sets. This signifies high accelerated expansion is supported by BAO method than DA method in the interval $z_2 < z < z_3$. The opposite happens the rest of the $z$ values.
\begin{figure}[h!]
\begin{center}
    $~~~~~~~~~~~~~Fig.5(a)~~~~~~~~~~~~~~~~~~~~~~~~~~~~~~~~Fig.5(b)~~~~~~~~~~~~~~~~~~~~~~~~~~~~~~~~~~Fig.5(c)~~~~~~~~~~$\\    
\includegraphics[height=2.1in,width=2.3in]{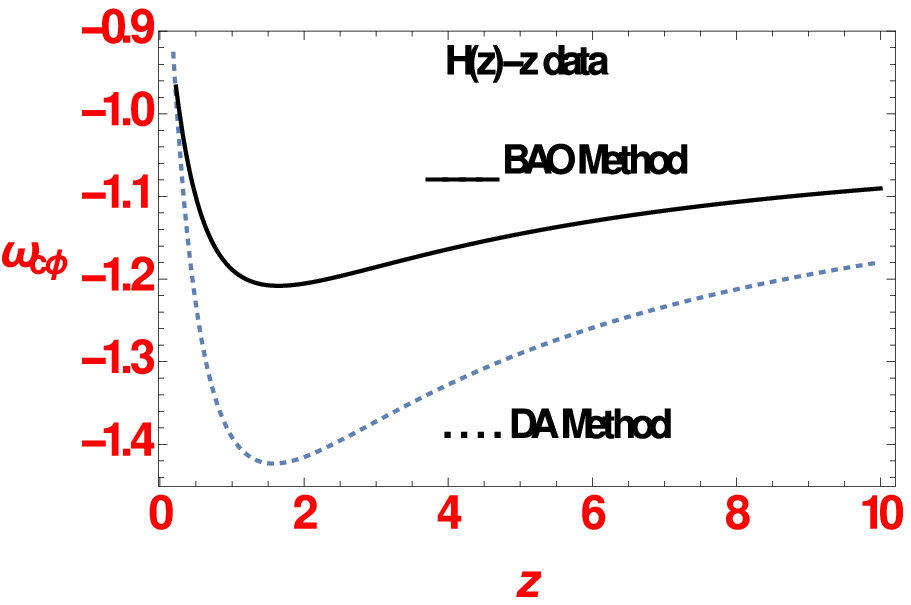}\includegraphics[height=2.1in,width=2.3in]{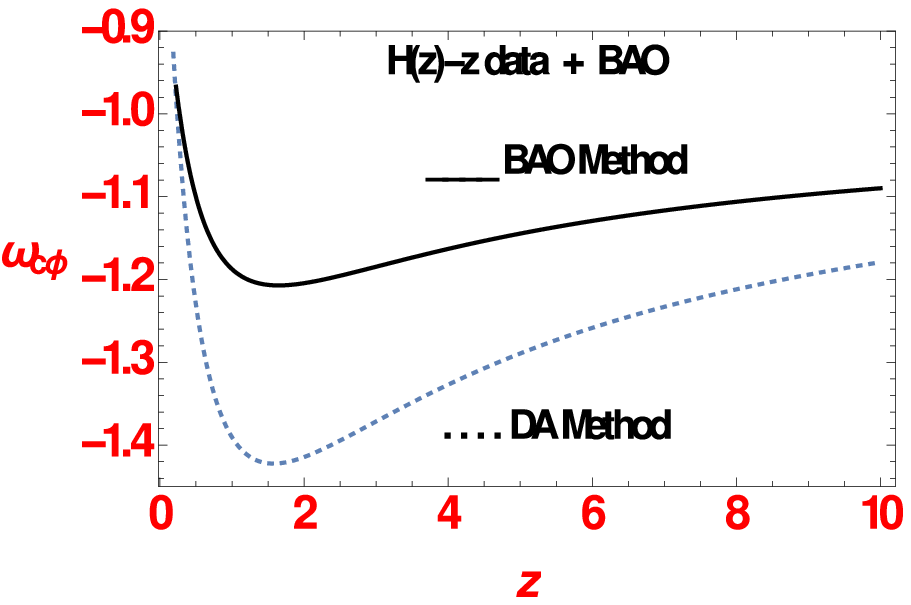}\includegraphics[height=2.1in,width=2.3in]{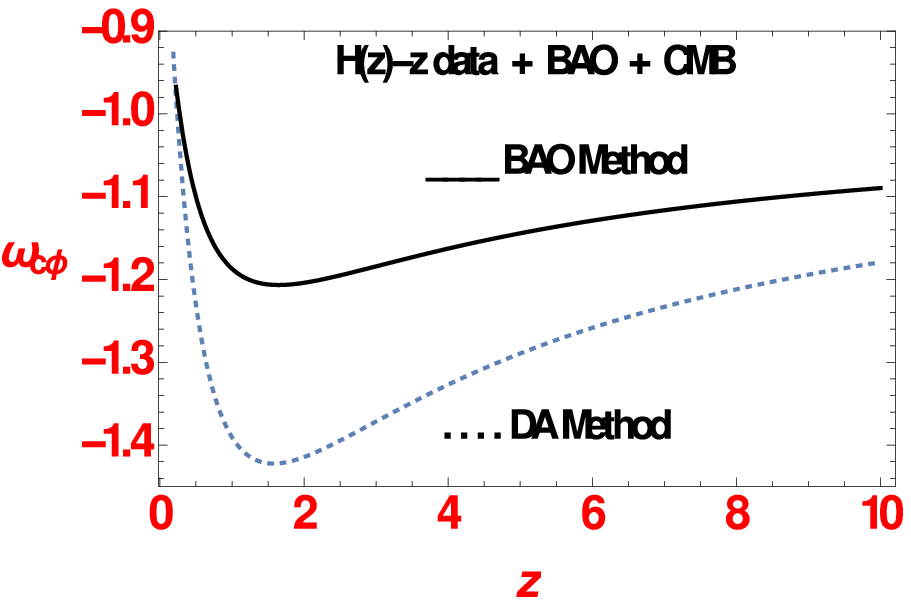}\\
Fig.5(a)-(c) : $\omega_{c\phi}$ vs $z$ graphs using $H(z)-z$ data, $H(z)-z$ data + BAO and $H(z)-z$ data with BAO +CMB where dotted lines shows DA method and solid line shows BAO method.\\
\end{center}
\end{figure}

In figure 5(a)-(c), we plot $\omega_{c\phi}$ vs z. The whole curve stays in negative zone/ fourth quadrant of $z-\omega_{c\phi}$ plane. The value of $\omega_{c\phi}$ is decreasing for a region of high $z$. As $z$ turns low, $\omega_{c\phi}$ starts to increase and its value becomes almost equal to -1 at a little past or a small neighbourhood of present time or $z = 0$. Tendency of $\omega_{c\phi}$ points that a future deceleration might be possible where the cosmos will transit from phantom to quintessence era again. This result does match with the work of Mamon, A. A. (2018). We find several works Shafieloo, A. et al. (2009), Li, Z. (2010), Magana, J. et al. (2014) where authors have presented that cosmic acceleration is presently witnessing its slowing down. Distinct methods have been used to establish the results. In our model we also find that universe has accelerated its expansion untill recent past. However, in present epoch and preferably in near future, the cosmos is likely to slow down its acceleration. This result is consistent with the above mentioned articles.
\begin{figure}[ht]
\begin{center}
$~~~~~~~~~~~~~Fig.6(a)~~~~~~~~~~~~~~~~~~~~~~~~~~~~~~~~~Fig.6(b)~~~~~~~~~~~~~~~~~~~~~~~~~~~~~~~~~~~~~~Fig.6(c)~~~~~~~~~~$\\    
\includegraphics[height=2.1in,width=2.3in]{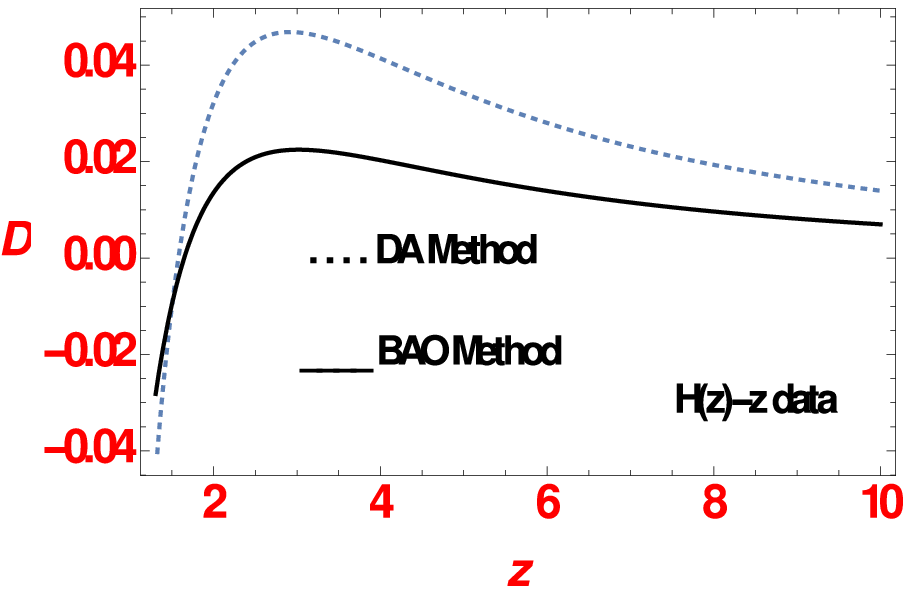}
\includegraphics[height=2.1in,width=2.3in]{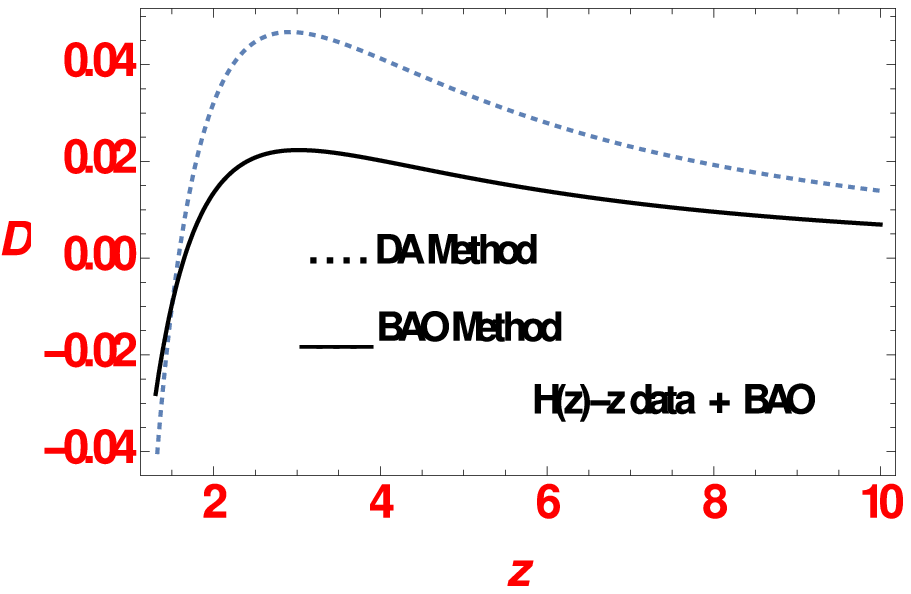}
\includegraphics[height=2.1in,width=2.3in]{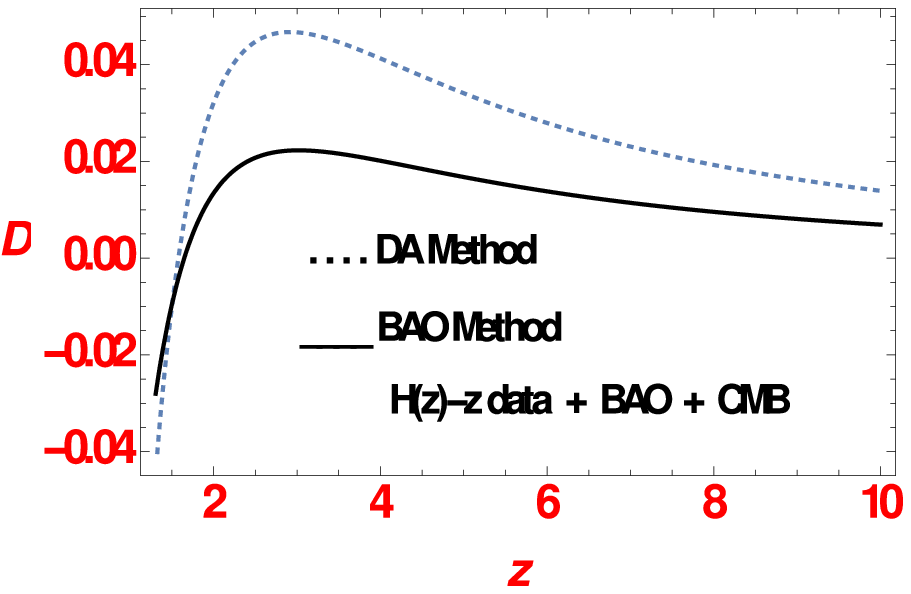}\\    
Fig.6(a)-(c) : Using $H(z)-z$ data, $D =\frac{d\omega_{c\phi}}{dz}$ vs $z$ graphs are drawn for $H(z)-z$ data, $H(z)-z$ data + BAO and $H(z)-z$ data + BAO + CMB respectively, where dotted lines describes DA method and solid line describes BAO method.\\
\end{center}
\end{figure}

Fig 6(a)-(c) are plots of $\frac{d\omega_{c\phi}}{dz}$ vs $z$. The rate of changes of $\omega_{c\phi}$ with respect to $z$ increases in high $z$ and then it falls near the present time.

\section{Brief Discussions and Conclusions :}
This article comprises of the construction of a redshift dependent model of dark energy and this model's behaviour under the constraints given by two particular redshift-Hubble parameter data sets. We have started with a cosmological model which is mainly governed by two independent components of Einstein's field equations for FLRW metric and equation of continuity for energy and matter. We have noticed that only three among these four governing equations can be independent of each other. But we were to solve four different quantities. This is why the requirement to consider a fourth is followed. This we have done by a process which gave birth of a new equation of state for the corresponding dark energy present in the current cosmos. As we have introduced a new dark energy representative, we require to specify the values of its different parameters. To do so we have motivated ourselves to constrain the parameters for two $H(z)-z$ data sets : derived from namely the differential ages method and Baryonic Acoustic Oscillation method. Firstly we plot these two data with each other and observe that the $H(z)$ graph is almost increasing with respect to corresponding $z$. The interesting part of this graph is that the higher values of $H(z)$ for low redshift can be seen in BAO case rather than DA method. The similar opposite phenomena happens for higher values of both redshift and $H(z)$ in DA method compared to BAO case. Next, we locate the best fit values of two parameters of our model under the data sets obtained by DA method and BAO method (along with BAO scaling and CMB constraints). We have plotted the $1\sigma$, $2\sigma$ and $3\sigma$ confidence contours for both data sets. We have observed that the contours are elliptic type, the semi major axis of which is inclined with a slope greater than one right angle. Confidence contours for DA method is almost a superset of that for BAO method. So BAO method constrains the model more than DA method does. Increment in the values of $k_1$ and $k_2$ increases the span of $\lambda_1$ and $\lambda_2$ causes the increment of the span of $\lambda_1$-$\lambda_2$ confidence contours in different ways. We have noted down the best its of the parameters $\lambda_1$ and $\lambda_2$ along with the span of the confidence contours in different tables. Fractional dimensionless densities for our model lies in the interval $[0,1]$ and fractional density for dark energy increases with time. On the other hand, deceleration parameter's value decreases with time.

The interesting result is found when we check the variation of the equation of state parameter with redshift. EoS parameter decreases with time and then increases again to become equal to almost $-1$ at the past neighbourhood of present time. Tendency of $\omega(z)$ shows a possibility future deceleration. From our model, we can theoretically construct the algebraic structure of the deceleration parameter, fraction dimensionless density, $\frac{d\omega_{c\phi}(z)}{dz}$ etc. We have plotted them as well to understand the deeper insight.

Variations of fractional dimensional densities show quite interesting phenomena. We observe the fractional density of dark energy to start almost from zero at high $z$ and to grow gradually. The same parameter for matter shows completely the opposite behaviour. For low redshift, the fractional density for dark energy grows high and almost becomes asymptotic to unity. This matches with a delayed decay of dark matter into dark energy with time.

Study of deceleration parameter vs redshift does not show any transition from deceleration to acceleration or converse. For present time neighbourhood $q$ falls abruptly. Variation of $\omega(z)$ shows that at present epoch $\omega(z)$ is converging to -1, especially when $H(z)-z$ data + BAO + CMB is applied as constraining tool.

To conclude in brief, our model which is of inverse quadratic nature fits with $H(z)$ vs $z$ data sets derived with the help of different ages method and Baryonic Acoustic Oscillation method by giving the best fits of the ($\lambda_1, \lambda_2$) type as (0.489817, -2.75048) and (0.118341, -0.289766) respectively. Both the data sets indicates to a $\omega \sim -1$ cosmology in $z=0$ epoch. DA does it with a prompt jump than a slower slope of BAO. Probable decay from matter to energy in late time universe is noted.

{\bf Acknowledgments:}
This research is supported by the project grant of Government of West Bengal, Department of Higher Education, Science and Technology and Biotechnology (File no:- $ST/P/S\&T/16G-19/2017$). PB thanks Department of Higher Education, Science \& Technology and Biotechnology, West Bengal for Swami Vivekananda Merit-Cum-Means Scholarship. RB thanks IUCAA, Pune for Visiting Associateship.


{\bf References :}
    
Alam, S., et al. :- {\it ``The clustering of galaxies in the completed SDSS-III Baryon Oscillation Spectroscopic Survey: cosmological analysis of the $DR12$ galaxy sample''} , (2016), [arXiv:1607.03155]

Alam, U., Sahni, V., Saini, T. D. \& Starobinski, A. A. :- {\it ``Is there supernova evidence for dark energy metamorphosis?''} , {\it Mon. Not. R. Astron. Soc.}, {\bf 354} (2004a) 275 [arXiv:0311364].\

Alam, U., Sahni, V. \& Starobinski, A. A. :- {\it ``The case for dynamical dark energy revisited''} , {\it JCAP}, {\bf 0406}, (2004b) 008 [arXiv:0403687v2].\

Alcaniz, J. S. \& Lima, J. A. S. :- {\it ``Dark Energy And The Epoch Of Galaxy Formation"} {\it ApJ}, {\bf 550}, L133 (2001).\

Anderson, L. et al. [BOSS Collaboration] :- {\it ``The clustering of galaxies in the SDSS-III Baryon Oscillation Spectroscopic Survey: baryon acoustic oscillations in the Data Releases 10 and 11 Galaxy samples.''} , {\it  Mon. Not. R. Astron. Soc}, {\bf 441(1)} (2014) 24 [arXiv:1312.4877]. \

Barboza Jr., E. M. \& Alcaniz, J. S. :- {\it ``A parametric model for dark energy''} , {\it Phys. Lett. B}, {\bf 666}, (2008) 415 [arXiv:0805.1713v1].\

Bassett, B. A. \& Hlozek, R. :- {\it ``Baryon Acoustic Oscillations"} , {\it Dark Energy, Ed. P. Ruiz-Lapuente}, (2010) [arXiv:0910.5224v1].\

Bautista, J. E. et al. :- {\it ``Measurement of baryon acoustic oscillation correlations at $z = 2.3$ with SDSSDR12 $L_{y\alpha}$-Forests''} ,{\it Astron. Astrophys.},{\bf 603}, (2017) 23 [arXiv:1702.00176].\

Bernardis, P. de et al.[Boomerang Collaboration] :- {\it A Flat Universe from High-Resolution Maps of the Cosmic Microwave Background Radiation"} ,{\it Nature} {\bf 404}, (2000) 955 [arXiv:0004404v1].\

Biswas, R. \& Debnath, U. :- {\it ``Constraining redshift parametrization parameters of dark energy: loop quantum gravity as background''} , {\it Eur.Phys.J.C}, {\bf 73} (2013) no.5, 2424.\

Biswas, P. \& Biswas, R. :- {\it ``Evolution of universe as a homogeneous system: Changes of scale factors with different dark energy Equation of States''} , {\it Modern Physics Letters A}, {\bf 33}, No. 19, (2018) 1850106 [arXiv:1710.06307].\

Biswas, P. \& Biswas, R. :- {\it ``Interacting Models of Generalized Chaplygin Gas and Modified Chaplygin Gas with Barotropic Fluid''} , {\it Modern Physics Letters A}, {\bf 34} 9(2019a)1950064[arXiv:1805.03962].\

Biswas, P. \& Biswas, R :- {\it ``Barboza-Alcaniz Equation of State Parametrization : Constraining the Parameters in Different Gravity Theories''} , {\it Modern Physics Letters A}, {\bf 34} 21(2019b)1950163[arXiv:1807.10608].\

Blake, C. et al. :- {\it ``The WiggleZ Dark Energy Survey: joint measurements of the expansion and growth history at $z \leq 1$.''} , {\it Mon. Not. R. Astron. Soc}, {\bf 425} (2012) 405 [arXiv:1204.3674].\

Busca,  N. G. et al. :- {\it ``Baryon acoustic oscillations in the Ly alpha forest of BOSS quasars''} , {\it Astron. \& Astrophys.}, {\bf 552} (2013) 18 [arXiv:1211.2616].\

Chakraborty, S. et al. :- {\it ``A third alternative to explain recent observations: Future deceleration"} , {\it Phys. Lett. B}, {\bf 738} (2014) 424 [arXiv:1411.0941].\

Chevallier, M., Polarski, D. :- {\it ``Acclerating Universes with Scaling Dark Matter ''} , {\it Int. J. Mod. Phys. D}, {\bf 10}, (2001) 213 [arXiv:gr-qc/0009008].\

Chuang, C-H. et al. :- {\it ``The clustering of galaxies in the SDSS-III Baryon Oscillation Spectroscopic Survey : single-probe measurements and the strong power of $f(z)\sigma8(z)$ on constraining dark energy''} , {\it Mon. Not.Roy. Astron. Soc.}, {\bf 433}, (2013). 3559-3571 [arXiv:1303.4486].\

Chuang, C.H. \& Wang, Y. :-{\it ``Modeling the Anisotropic Two-Point Galaxy Correlation Function on SmallScales and Improved Measurements of $H(z)$, $DA(z)$, and $f(z)$ $\sigma 8(z)$ from the Sloan Digital Sky Survey DR7 Luminous Red Galaxies''} , {\it Mon. Not. Roy. Astron. Soc.}, {\bf 435}, (2013) 255-262 [arXiv:1209.0210].\

Cole, S. et al. :-{\it ``The 2dF Galaxy Redshift Survey: power-spectrum analysis of the final data set and cosmological implications''} , {\it  Mon. Not. Roy. Astron. Soc.}, {\bf 362} (2005) 505 [arXiv:0501174].\

Copeland, E. J., Sami, M. \& Tsujikawa, S. :- {\it ``Dynamics of dark energy"} , {\it Int. J. Mod. Phys. D} , {\bf 15} (2006) 1753 [arXiv:0603057v3].\

Delubac , T. et al. [BOSS Collaboration], {\it ``Baryon acoustic oscillations in the $L_y$ forest of BOSS DR11 quasars''} , {\it Astron. Astrophys.}, {\bf 574} (2015) A59 [arXiv:1404.1801].\

Doran, M., Stern. S, \& Thommes, E. :- {\it ``Baryon acoustic oscillations and dynamical dark energy"} , {\it JCAP}, {\bf 0704}, (2007) 015 [arXiv:0609075].\

Dunlop, J. et al. :- {\it ``A $3.5$-Gyr-old galaxy at redshift 1.55"}, {\it Nature}, {\bf 381}, 581 (1996).\

Efstathiou, G. :- {\it ``Constraining the equation of state of the Universe from Distant Type Ia Supernovae and Cosmic Microwave Background Anisotropies"} , {\it Mon. Not. R. Astron. Soc.}, {\bf 310}, (1999) 842 [arXiv:9904356v1].\

Efstathiou, G. \& Bond J. R. :- {\it ``Cosmic confusion: degeneracies among cosmological parameters derived from measurements of microwave background anisotropies"} , {\it MNRAS}, {\bf 304}, (1999) 75 [arXiv:9807103].\

Einstein, A. :- {\it ``Cosmological Considerations in the General
Theory of Relativity."}; {\it Sitzungsber. Preuss. Akad. Wiss. Berlin (Math. Phys.),} {\bf 142} (1917).\

Eisenstein, D. J. et al. (SDSS collaboration):- {\it ``Detection of the baryon acoustic peak in the large-scale correlation function of SDSS luminous red galaxies''} , {\it Astrophys. J.}, {\bf 633} (2005) 560 [arXiv:0501171].\

Elgaroy, O. \& Multamaki, T. :- {\it ``On using the cosmic microwave background shift parameter in tests of models of dark energy"} , {\it Astron. Astrophys}, {\bf 471} (2007) 65E [arXiv:0702343].\

Feng, C., J., Shen, X. -Y., Li, P. \& Li, X. -Z. :- {\it A New Class of Parametrization for Dark Energy without Divergence" }, {\it JCAP}, {\bf 1209}, (2012) 023,  [arXiv:1206.0063].\

Font-Ribera, A., et al. :- {\it ``Quasar-Lyman $\alpha$ Forest Cross-Correlation from BOSS DR11 : Baryon Acoustic Oscillations''} , {\it JCAP}, {\bf 1405}, (2014) 027 [arXiv:1311.1767].\

Friedmann, A. :- {\it ``On the curvature of space."} ; {\it  Z. Phys.} {\bf 10}, (1922) 377.\

Gazta$\tilde{n}aga$,  E., Cabr$e\backprime$, A. \& Hui, L.  :- {\it ``Clustering of luminous red galaxies - IV. Baryonacoustic peak in the line-of-sight direction and a direct measurement of $H(z)$''} , {\it Mon. Not. R.Astron. Soc.}, {\bf 399} (2009) 1663 [arXiv:0807.3551]\

Hanany, S. et al. :- {\it ``MAXIMA-1: A Measurement of the Cosmic Microwave Background Anisotropy on angular scales of 10 arcminutes to 5 degrees"} , Astrophys. J. {\bf 545}, (2000) L5 [arXiv:0005123].\

Hannestad, S. \& Mörtsell, E. :- {\it ``Cosmological constraints on the dark energy equation of state and its evolution"}, {\it JCAP}, {\bf 0409}, (2004) 001, [arXiv:0407259].\

Hubble, E. :- {\it ``A relation between distance and radial velocity among extra–galactic nebulae."} {\it Proc. Nat. Acad. Sci.} {\bf 15}, (1929) 168.\

Jassal, H. K., Bagla, J. S., Padmanabhan, T. :- {\it ``WMAP constraints on low redshift evolution of dark energy"} , {\it Mon. Not. R. Astron. Soc.}, {\bf 356},  (2005) L11 333  [arXiv:astro-ph/0404378].\

Jimenez, R. et al. :- {\it ``Premature dismissal of high-redshift elliptical galaxies"}, {\it Mon. Not. R. Astron. Soc.}, {\bf 305}, (1999) L16–L20 , [arXiv:9812222].\

Jimenez, R. et al. :- {\it ``Constraints on the equation of state of dark energy and the Hubble constant from stellar ages and the cosmic microwave background"}, {\it Astrophys. J}, {\bf 593}, (2003) 622–629, [arXiv:0302560].\

Jimenez, R. \& Loeb, A. :- {\it ``Constraining cosmological parameters based on relative galaxy ages''} , {\it Astrophys. J.}, {\bf 573} (2002) 37 [arXiv:astro-ph/0106145].\

Lee, S. :- {\it ``Constraints on the dark energy equation of state from the separation of CMB peaks and the evolution of alpha "}, {\it Phys. Rev. D}, {\bf 71}, (2005) 123528 [arXiv:0504650].\

Lemaitre, G. :- {\it ``Un Univers homogène de masse constante et de rayon croissant rendant compte de la vitesse radiale des nébuleuses extra-galactiques.}; {\it Annales de la Société Scientifique de Bruxelles} {\bf 47}, (1927) 49.\

Li, Z. et al. :- {\it ``Probing the course of cosmic expansion with a combination of observational data"} , {\it JCAP}, {\bf 11} (2010) 31 [arXiv:1011.2036].\

Lima, J. A. S. \&  Alcaniz, J. S. :- {\it ``Flat FRW Cosmologies with Adiabatic Matter Creation: Kinematic tests''} , {\it Astron. Astrophys.}, {\bf 348}, (1999) 1 [arxiv:9902337].\

Linder, E. V. :- {\it ``Exploring the Expansion History of the Universe"} , {\it Phys. Rev. Lett.}, {\bf 90} (2003) 091301 [arXiv:0208512v1].\

Linder, E. V. :- {\it ``Cosmic growth history and expansion history"} , {\it Phys. Rev. D} {\bf 72}, (2005) 043529 [arXiv:0507263].\

Macaulay, E. et. al. (DES collaboration) :- {\it ``First Cosmological Results using Type Ia Supernovae from the Dark Energy Survey: Measurement of the Hubble Constant''} , [arXiv:1811.02376].\

Magana, J. et al. :- {\it ``Cosmic slowing down of acceleration for several dark energy parametrizations"} , {\it JCAP}, {\bf 017} (2014) 1410 [arXiv:1407.1632].\

Mathews, G. J., Lan,  N. Q. \& Kolda, C. :- {\it ``Late decaying dark matter, bulk viscosity, and the cosmic acceleration"} , {\it Phys. Rev. D}, {\bf 78}, (2008) 043525 [arXiv:0801.0853].\

McCarthy, P. J. et al. :- {\it ``Evolved galaxies at $z >1.5$ fromthe Gemini Deep Deep Survey: The formation epoch of massive stellar systems"}, {\it Astrophys. J}, {\bf 614} ,(2004) L9–L12, [arXiv:0408367].\

Moresco, M. et al. :- {\it ``Improved constraints on the expansion rate of the Universe up to $z \sim 1.1$ from the spectroscopic evolution of cosmic chronometers''} , {\it JCAP}, {\bf 08} (2012a) 006 [arXiv:1201.3609].\

Moresco, M. et al. :- {\it ``New constraints on cosmological parameters and neutrino properties using the expansion rate of the Universe to $z \sim 1.75$''} , {\it JCAP}, {\bf 07} (2012b) 053 [arXiv:1201.6658].\

Moresco, M. :- {\it ``Raising the bar: new constraints on the Hubble parameter with cosmic chronometers at $z \sim 2$''} , {\it Mon. Not. Roy. Astron. Soc.}, {\bf 450} (2015) L16 [arXiv:1503.01116]. \

Oka, A. et al. :- {\it ``Simultaneous constraints on the growth of structure and cosmic expansion from the multipole power spectra of the SDSS DR7 LRG sample''} , {\it Mon. Not. Roy. Astron. Soc.}, {\bf 439}, (2014) 2515-2530 [arXiv: 1310.2820].\

Ozer, M. \& Taha, M. O. :- {\it ``A possible solution to the main cosmological problems''} , {\it Phys. Lett. B}, {\bf 171} (1986) 363.\

Pan, S. \& Chakraborty, S. :- {\it ``Will there be again a transition from acceleration to deceleration in course of the dark energy evolution of the universe?"} , {\it Eur. Phys. J. C}, {\bf 73} (2013) 2575 [arXiv:1303.5602].\

Pan, Z., Knox, L., Mulroe, B. \& Narimani, A. :- {\it ``Cosmic Microwave Background Acoustic Peak Locations"} , {\it MNRAS}, {\bf 459} (2016) 2515 [arXiv:1603.03091v2].\

Peebles, P. J. E. \& Ratra, B. :- {``The Cosmological Constant and Dark Energy''} , {\it Rev. Mod. Phys.}, {\bf 75} (2003) 559 [arXiv:0207347v2].\

Perlmutter, S. et al. (Supernova Cosmology Project collaboration):- {\it `` Measurements of $\Omega$ and $\Lambda$ from 42 high redshift supernovae"} {\it  Astrophys. J.} {\bf 517} (1999) 565 [astro-ph/9812133].\

Ratra, B. \& Peebles,  P. J. E. :- {\it ``Cosmological consequences of a rolling homogeneous scalar field"} , {\it Phys. Rev. D} {\bf 37}, (1988) 3406.\

Ratsimbazafy, A. L. et al. :- {\it ``Age–dating Luminous Red Galaxies observed with the Southern African Large Telescope''} , {\it Mon. Not. Roy. Astron. Soc.}, {\bf 467} (2017) 3239 [arXiv:1702.00418].\

Riemer-Sorensen, S. et al. :- {\it ``Simultaneous constraints on the number and mass of relativistic species''}  , {\it Astrophys. J.}, {\bf 763} (2013) 89 [arXiv:1210.2131].\

Riess, A. G. et al. (Supernova Search Team collaboration) :- {\it `` Observational evidence from supernovae for an accelerating universe and a cosmological constant"} {\it  Astron. J.} {\bf 116} (1998) 1009 [astro-ph/9805201].\

Sahni, V. \& Starobinsky, A. A. :- {\it ``The Case for a Positive Cosmological Lambda-term''} , {\it Int. J. Mod. Phys. D}, {\bf 9} (2000) 373 [arXiv:9904398v2].\

Sendra, I. \& Lazkoz, R. :- {\it ``SN and BAO constraints on (new) polynomial dark energy parametrizations: current results and forecasts" }, {\it Mon. Not. Roy. Astron. Soc.}, {\bf 422}, (2012) 776 [arXiv:1105.4943].\

Shafieloo, A. et al. :- {\it ``Is cosmic acceleration slowing down?"}, {\it Phys. Rev. D}, {\bf 80}, (2009) 101301(R) [arXiv:0903.5141].\\ 

Silva, R., Alcaniz J. S. \& Lima, J. A. S. :- {it ``On the thermodynamics of dark energy"} , {\it Int. J. Mod. Phys. D} {\bf 16}, (2007) 469.\

Simon, J. et al. :- {\it ``Constraints on the redshift dependence of the dark energy potential''} , {\it Phys. Rev. D}, {\bf 71} 123001 (2005) [arXiv:0412269].\

Stern, D. et al. :- {\it ``Cosmic chronometers: constraining the equation of state of dark energy. I: $H(z)$ measurements''} , {\it JCAP}, {\bf 1002} (2010) 008 [arXiv:0907.3149].\

Stockton, A. :- {\it ``The Oldest Stellar Populations at $z \sim 1.5$"}, to be published in Astrophysical Ages and Time Scales, ASP Conference Series, (2001) [arXiv:0104191].\

Turner, M. S., Steigman, G. \&  Krauss, L. M. :- {\it ``Flatness of the Universe: Reconciling Theoretical Prejudices with Observational Data"} , {\it Phys. Rev.Lett.}, {\bf 52}, (1984) 2090.\

Valent, A. G. \& Amendola, L. :- {\it ``$H_0$ from cosmic chronometers and Type Ia supernovae, with Gaussian Processes and the novel Weighted Polynomial Regression method''} , {\it JCAP}, {\bf 1804} (2018) 051, [arXiv:2.01505v3].\

Wang, X. et al. :- {\it `` Observational constraints on cosmic neutrinos and dark energy revisited"}, {\it JCAP} {\bf 11} (2012) 018 [arXiv:1210.2136].\

Wang, Y. et al. :- {\it ``The clustering of galaxies in the completed SDSS-III BaryonOscillation SpectroscopicSurvey: tomographic BAO analysis of DR12 combined sample configuration space} , {\it Mon. Not. Roy.Astron. Soc.}, {\bf 469}, (2017) 3762–3774 [arXiv:1607.03154].\

Weller, J. \& Albrecht, A. :- {\it ``Future supernova observations as a probe of dark energy" }, {\it Phys. Rev. D}, {\bf 65}, (2002) 103512 [arXiv:0106079].

Zeldovich, Y. B. :-{\it `` Special Issue: the Cosmological Constant and the Theory of Elementary Particles."}; {\it Soviet Physics Uspekhi} {\bf 11}, (1968) 381.\

Zhang, C. et al.  :- {\it ``Four new observational $H(z)$ data from luminous red galaxies in the Sloan Digital Sky Survey data release seven''} , {\it Res. Astron. Astrophys.}, {\bf 14}, 1221 (2014) [arXiv:1207.4541].\

Zhao, G. B. et al. :- {\it ``Examining the evidence for dynamical dark energy''} , {\it Phys. Rev. Lett.}, {\bf 109} (2012) 171301 [arXiv:1207.3804].\
\end{document}